\documentclass[11pt,a4paper]{article} 
\usepackage{jheppub}
\pdfoutput=1
\usepackage{epsfig}
\usepackage{graphicx}
\usepackage{amsmath}
\usepackage{amsfonts}
\usepackage{amssymb}

\usepackage{color}

\def\beq{\begin{equation}}
\def\eeq{\end{equation}}
\def\beqa{\begin{eqnarray}}
\def\eeqa{\end{eqnarray}}
\newcommand{\nn}{\nonumber}

\def\eqn#1{eq.~(\ref{#1})}
\def\eqns#1#2{eqs.~(\ref{#1}) and~(\ref{#2})}
\def\eqnss#1#2{eqs.~(\ref{#1})-(\ref{#2})}
\def\Eqn#1{Eq.~(\ref{#1})}
\newcommand\Eqns[2]{Eqs.\,(\ref{#1}) and~(\ref{#2})}

\def\fig#1{fig.~{\ref{#1}}}
\def\sec#1{sec.~{\ref{#1}}}
\def\app#1{app.~\ref{#1}}

\def\bsp#1\esp{\begin{split}#1\end{split}}

\newcommand{\eps}{\epsilon}

\newcommand{\ord}{{\cal O}}
\def\cM{{\cal M}}
\def\cR{{\cal R}}
\newcommand{\RE}{{\rm Re}}

\def\ifm{\ifmmode}

\def \Tr {\mbox{Tr\,}}

\def\dsqcup{\sqcup\mathchoice{\mkern-7mu}{\mkern-7mu}{\mkern-3.2mu}{\mkern-3.8mu}\sqcup}

\def\cg{\kappa_\Gamma}
\newcommand\sss{\scriptscriptstyle}
\newcommand\as{\alpha_{\sss S}} 
\newcommand\gs{g_{\sss S}}

\def \al #1 {\frac {\as({#1})}{\pi} }
\def \ds #1 {\ooalign{$\hfil/\hfil$\crcr$#1$}}

\def\msb{\ifm \overline{\rm MS}\,\, \else $\overline{\rm MS}\,\, $\fi}

\def\eps{\varepsilon}

\allowdisplaybreaks[1]

\preprint{~}

\title{One-loop impact factor for the emission of two gluons}

\author[a,b]{Marc Canay}
\author[a,1]{Vittorio Del Duca}
\note{On leave from INFN, Laboratori Nazionali di Frascati, Italy.}
\affiliation[a]{Institute for Theoretical Physics, ETH Z\"urich, 8093 Z\"urich, Switzerland.}
\affiliation[b]{Institut de Physique Th\'eorique, CEA, CNRS, Universit\'e Paris--Saclay,
  F--91191 Gif-sur-Yvette cedex, France.}

\emailAdd{marc.canay@ipht.fr}
\emailAdd{delducav@itp.phys.ethz.ch}

\abstract{We consider one-loop five-point QCD amplitudes in next-to-multi-Regge kinematics, and evaluate
the one-loop impact factor for the emission of two gluons.
This is the last ingredient which is necessary to evaluate the gluon-jet impact factor at NNLO accuracy in $\as$.
It is also the first instance in which loop-level QCD amplitudes are evaluated in next-to-multi-Regge kinematics,
which requires to apply a different reggeisation ansatz to each colour-ordered amplitude.}


\begin{document}

\maketitle

\section{Introduction}
\label{sec:intro}

The Regge limit of $2\to 2$ scattering amplitudes of massless gauge theories, like Yang-Mills theories or ${\cal N}=4$ Super Yang-Mills (SYM),
has been studied since the pioneering work of Lipatov on the gluon Reggeisation~\cite{Lipatov:1976zz}, and the more so
in the last decade, after realising that amplitudes in the Regge limit are endowed with a rich mathematical 
structure~\cite{Dixon:2012yy,DelDuca:2013lma,DelDuca:2016lad,DelDuca:2017peo,DelDuca:2019tur,Caron-Huot:2020grv}.

\begin{figure}
  \centerline{\includegraphics[width=0.25\columnwidth]{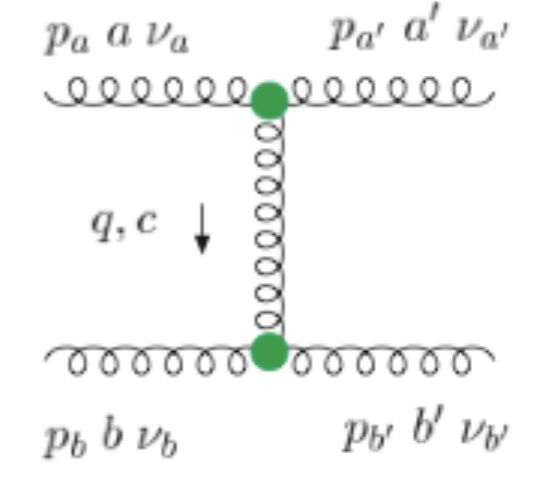} }
  \caption{Amplitude for $gg\rightarrow gg$ in the Regge limit. The green blobs represent the impact factors.}
\label{fig:llbfkl}
\end{figure}

In the Regge limit, in which the squared centre-of-mass energy $s$ is much larger than the momentum transfer $t$, $s\gg |t|$, 
$2\to 2$ scattering amplitudes are dominated by gluon exchange in the $t$ channel. Contributions which do not feature gluon exchange in the $t$ channel
are power suppressed in $t/s$. In particular, at tree level we can write the $2\to 2$ amplitudes in a factorised way. For example, 
the tree amplitude for gluon-gluon scattering $g_a\, g_b\to g_{a'}\,g_{b'}$ at fixed helicities may be written as \cite{Lipatov:1976zz,Kuraev:1976ge}, 
\begin{equation}
\cM^{(0)}_{gg\to gg} = 
\left[\gs (F^{a'})_{ac}\, C^{g(0)}(p_a^{\nu_a};p_{a'}^{\nu_{a'}}) \right]
{s\over t} \left[\gs (F^{b'})_{cb}\, C^{g(0)}(p_b^{\nu_b};p_{b'}^{\nu_{b'}}) 
\right]\, ,\label{elas}
\end{equation}
with $s=(p_a+p_b)^2$, $q = p_{b'} + p_b$ and $t=q^2  \simeq -|q_\perp|^2$, \fig{fig:llbfkl}, and $(F^c)_{ab} = i\sqrt{2} f^{acb}$, 
and where the superscripts $\nu$ label the helicities\footnote{We take all the momenta as outgoing, so the helicities for incoming partons are reversed, see \app{sec:appa}.}.
As it is apparent from the colour coefficient  $(F^{a'})_{ac} (F^{b'})_{cb}$, in \eqn{elas} only the antisymmetric octet ${\bf 8}_a$
is exchanged in the $t$ channel.
Finally, \eqn{elas} yields the four-gluon amplitude in terms of one colour-ordered amplitude, in agreement with the Bern-Carrasco-Johansson (BCJ) relations~\cite{Bern:2008qj}, which fix at $(n-3)!$ the number of independent colour-ordered amplitudes.

Since the four-gluon amplitude is a maximally helicity-violating (MHV) amplitude, \eqn{elas} describes $\binom{4}{2} = 6$ 
helicity configurations. However, at leading power in $t/s$, helicity is conserved along the $s$-channel direction in Minkowski space, so in \eqn{elas} four helicity configurations are leading, two for each tree impact factor, $g^*\, g \rightarrow g$, with $g^*$ an off-shell gluon~\cite{DelDuca:1995zy}\footnote{In the literature, the colour coefficient of the gluon-gluon scattering amplitude in the Regge limit is usually written as $f^{aa'c} f^{bb'c}$. In \eqn{elas}, we follow the cyclic ordering of the gluons in the amplitude, obtaining a relative minus sign, which is absorbed into the definition of $C^{g(0)}(p_b^-;p_{b'}^+)$ in \eqn{centrc}. },
\begin{equation}
C^{g(0)}(p_a^-;p_{a'}^+) = 1\,, \qquad C^{g(0)}(p_b^-;p_{b'}^+) =
- {p_{b'\perp}^* \over p_{b'\perp}}\, .\label{centrc}
\end{equation} 
The impact factors transform under parity into their complex conjugates,
\begin{equation}
[C^{g}(p^\nu;p^{\nu'})]^* = C^{g}(p^{-\nu};p^{-\nu'})\, . 
\end{equation} 
The helicity-flip impact factor $C^{g(0)}(p^+;p'^+)$ and its parity conjugate $C^{g(0)}(p^-;p'^-)$ 
are power suppressed in $t/s$.

The tree amplitudes for quark-gluon or quark-quark scattering have the same form as \eqn{elas}, up to replacing one or both
gluon impact factors $C^{g(0)}$ (\ref{centrc}) with quark impact factors $C^{q(0)}$, and the colour factors $(F^c)_{ab}$ in the adjoint representation
with the colour factors $T^c_{ij}$ in the fundamental representation of SU(3), normalised as $\Tr(T^aT^b)= T_F \delta^{ab}$, with $T_F=1$.
So in the Regge limit, the $2\to 2$ scattering amplitudes factorise into gluon or quark impact factors
and a gluon propagator in the $t$ channel, and are uniquely determined by them.

The loop corrections to an amplitude introduce poles and branch cuts, which are dictated by the analytic
structure and constrained by the symmetries of the amplitude. 
In particular, in the Regge limit $s\simeq -u \gg -t$, $2\to 2$ scattering amplitudes are symmetric under $s\leftrightarrow u$ crossing,
thus we may consider amplitude states whose kinematic and colour coefficients have a definite signature under $s\leftrightarrow u$ crossing,
\begin{equation}
\cM_4^{(\pm)}(s,t) = \frac{ \cM_4(s,t) \pm \cM_4(u,t) }{2}\,,
\label{eq:sucross}
\end{equation}
with $u=-s-t\simeq -s$, such that $\cM_4^{(-)}(s,t)$ ($\cM_4^{(+)}(s,t)$) has kinematic and colour coefficients which are both odd (even) 
under $s\leftrightarrow u$ crossing.
Further, higher-order contributions to $g g \rightarrow g g$ scattering
in general involve additional colour structures, as dictated by the decomposition of the product ${\bf 8}_a \otimes {\bf 8}_a$
into irreducible representations,
\beq
  {\bf 8}_a \otimes {\bf 8}_a \, = \, \{ {\bf 1} \oplus {\bf 8}_s \oplus {\bf 27} \} 
  \oplus [ {\bf 8}_a \oplus {\bf 10} \oplus \overline{{\bf 10}} ]\ \, ,
\label{8*8}
\eeq
where in curly (square) brackets are the representations which are even (odd) under $s\leftrightarrow u$ crossing.

\subsection{The Regge limit at leading logarithmic accuracy}
\label{sec:reggell}

When loop corrections to the tree amplitude (\ref{elas}) are considered, 
it is found that at leading logarithmic (LL) accuracy in $\log(s/|t|)$, the four-gluon amplitude is given to all orders
in $\alpha_s$ by~\cite{Lipatov:1976zz,Kuraev:1976ge}
\begin{equation}
\left. \cM_{gg\to gg}\right|_{LL} =  \left[\gs (F^{a'})_{ac}\, C^{g(0)}(p_a^{\nu_a}; p_{a'}^{\nu_{a'}}) \right] {s\over t}
\left({s\over \tau}\right)^{\alpha(t)}\, \left[\gs (F^{b'})_{cb}\, C^{g(0)}(p_b^{\nu_b}; p_{b'}^{\nu_{b'}}) \right]\,,
\label{sud}
\end{equation}
with $\alpha(t)$ related to the loop transverse-momentum integration, and where $\tau > 0$ is a Regge factorisation scale.
$\tau$ is of order of $t$, and much smaller than $s$, however the precise definition of $\tau$ is immaterial to LL accuracy,
where one can suitably fix $\tau = - t$. 
Regulating $\alpha(t)$ in $d=4-2\epsilon$ dimensions, one obtains
\begin{equation}
\alpha(t) \equiv \frac{\gs^2}{(4\pi)^2} \alpha^{(1)}(t) = \frac{\as}{4\pi} N_c\, {2\over\epsilon} 
\left(\mu^2\over -t\right)^{\epsilon} \kappa_{\Gamma}\, ,\label{alph}
\end{equation}
with $N_c$ the number of colours, $\as = \gs^2/(4\pi)$ and 
\begin{equation}
\kappa_\Gamma = (4\pi)^\epsilon\, {\Gamma(1+\epsilon)\,
\Gamma^2(1-\epsilon)\over \Gamma(1-2\epsilon)}\, .\label{cgam}
\end{equation}
The prominent facts of eq.~(\ref{sud}) are that at LL accuracy the amplitude (\ref{sud}) is still real,
the one-loop result (\ref{alph}) exponentiates, and the antisymmetric octet ${\bf 8}_a$
is still the only colour representation exchanged in the $t$ channel. Thus, at LL accuracy,
\begin{equation}
\left. \cM_{gg\to gg}\right|_{LL} = \left. \cM^{(-)[8_a]}_{gg\to gg}\right|_{LL}\,, \qquad\qquad \left. \cM^{(+)}_{gg\to gg}\right|_{LL} = 0\,.
\end{equation}
$\alpha(t)$ is called Regge trajectory, and the exponentiation of $\log(s/|t|)$
in the one-loop result (\ref{alph}) is called gluon Reggeisation~\cite{Balitsky:1979ap}.
Because of \eqn{sud}, factorisation holds at LL accuracy just like at tree level, i.e. the 
amplitudes for quark-gluon or quark-quark scattering at LL accuracy
have the same form as \eqn{sud}, up to replacing one or both
colour and impact factors for gluons with the ones for quarks.

The gluon Reggeisation at LL accuracy constitutes
the backbone of the Balitsky-Fadin-Kuraev-Lipatov (BFKL) 
equation~\cite{Fadin:1975cb,Kuraev:1976ge,Kuraev:1977fs,Balitsky:1978ic},
which at $t=0$ describes the $s$-channel cut forward amplitude, which through the
optical theorem is equivalent to the squared amplitude integrated over all the allowed final states.
The BFKL equation sums the terms of ${\cal O}(\as^n \log^n(s/|t|))$ and
describes the evolution of the gluon ladder in transverse momentum space,
after the gluon rapidities have been integrated out.
Since in the BFKL equation also the real emissions are included,
in order to match the LL accuracy of the virtual corrections (\ref{sud}),
amplitudes with five or more gluons are taken in the
multi-Regge kinematics (MRK), which requires that the gluons
are strongly ordered in rapidity and have comparable transverse momentum, 
see \app{sec:appb}.
\begin{figure}
  \centerline{\includegraphics[width=0.25\columnwidth]{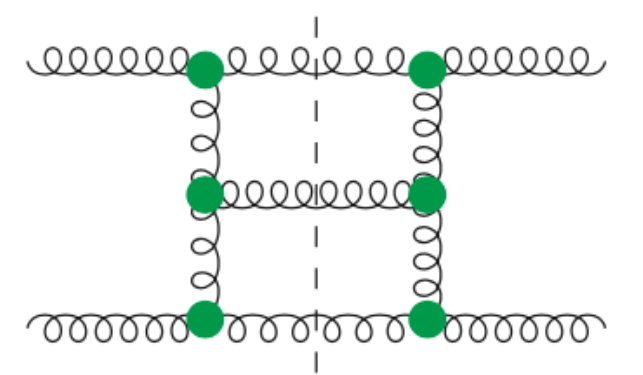} ~~~~~~
  \includegraphics[width=0.25\columnwidth]{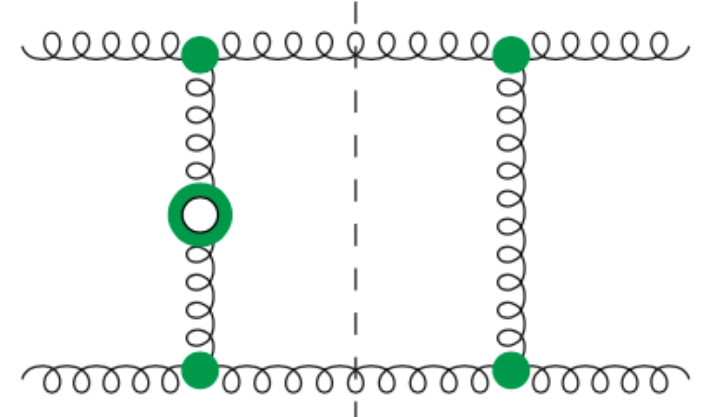}  }
  \caption{$(a)$ Tree five-gluon amplitude. The green blob along the gluon ladder represents the central-emission vertex.
  $(b)$ Regge trajectory in the one-loop four--gluon amplitude. The pierced blob represents the one-loop Regge trajectory.}
\label{fig:oneloopregge}
\end{figure}
The MRK rationale is that each gluon emitted along the ladder requires a factor of $\as$ and after
being integrated out in rapidity yields a factor of $\log(s/|t|)$, in such a way that each gluon
emitted along the ladder contributes a factor of ${\cal O}(\as \log(s/|t|))$.
Five-gluon amplitudes in MRK allow one to compute the emission of a gluon along the gluon ladder.
The factor associated to that gluon emission is called central-emission vertex, or Lipatov vertex~\cite{Lipatov:1976zz}.
The central-emission vertex constitutes the kernel of the BFKL equation.
Beyond five gluons, amplitudes in MRK assume an iterative structure, with a central-emission vertex
for each gluon emitted along the ladder. Thus, the building blocks of the BFKL equation are the
central-emission vertex at tree level, \fig{fig:oneloopregge}$(a)$,
and the Regge trajectory at one loop (\ref{alph}), \fig{fig:oneloopregge}$(b)$.
Finally, the impact factors (\ref{centrc}) are not part of the gluon ladder, but sit at its ends, and 
when they are squared the constitute the jet impact factors and contribute to any jet cross section~\cite{Mueller:1986ey,DelDuca:1993mn,Stirling:1994he,Andersen:2001kta,Andersen:2008gc,Andersen:2009nu,Andersen:2011hs} 
which is computed through the BFKL equation.

\subsection{The Regge limit at next-to-leading logarithmic accuracy}
\label{sec:reggenll}

\begin{figure}
  \centerline{\includegraphics[width=0.2\columnwidth]{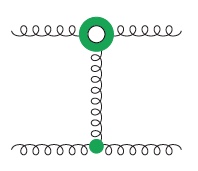} ~~~~
  \rotatebox[origin=c]{180}{\includegraphics[width=0.2\columnwidth]{1loopggIF.jpeg}  } }
  \caption{Impact factors in the one-loop four--gluon amplitude. The pierced blobs represent the one-loop impact factor $C^{g (1)}(t)$.}
\label{fig:oneloopif}
\end{figure}

At next-to-leading-logarithmic (NLL) accuracy, 
keeping the $s\leftrightarrow u$ crossing symmetry (\ref{eq:sucross}) into account, 
the exchange of one Reggeised gluon of \eqn{sud}
generalises to~\cite{Fadin:1993wh}
\beqa
\lefteqn{ \cM^{(-)[8_a]}_{gg\to gg} } \label{sudall}\\
&=& \frac1{2}  
\left[\gs (F^{a'})_{ac}\, C^{g}(p_a^{\nu_a};p_{a'}^{\nu_{a'}}) \right] {s\over t} 
\left[ \left({s\over \tau}\right)^{\alpha(t)} + \left({-s\over \tau}\right)^{\alpha(t)} \right]
\, \left[\gs (F^{b'})_{cb}\, C^{g}(p_b^{\nu_b};p_{b'}^{\nu_{b'}}) \right]\,, \nn
\eeqa
where the colour and kinematic parts of the amplitude are each odd under $s\leftrightarrow u$ crossing.
Further, by expanding the gluon Regge trajectory,
\begin{equation}
\alpha(t) = \frac{\as}{4\pi} \alpha^{(1)}(t) + 
\left(\frac{\as}{4\pi}\right)^2 \alpha^{(2)}(t) + \ord(\as^3)\,
,\label{alphb}
\end{equation}
with $\alpha^{(1)}(t)$ given in \eqn{alph}, and the helicity-conserving impact factor,
\begin{equation}
C^{g}(p_j^\mp;p_{j'}^\pm) = 
C^{g (0)} (p_j^\mp;p_{j'}^\pm)\left(1 + \frac{\as}{4\pi} C^{g (1)}_{-+}(t,\tau) + \ord(\as^2) \right)\, 
.\label{fullv}
\end{equation}
with $j = a, b$, see \fig{fig:oneloopif}, one can write the amplitude (\ref{sudall}) as a double expansion in the strong coupling $\as$
and in $\log(s/\tau)$, 
\beq
 \cM^{(-)[8_a]}_{gg\to gg} =  \cM^{(0)}_{gg\to gg}
\left( 1 + \sum_{\ell=1}^\infty \left(\frac{\as}{4\pi}\right)^\ell M^{(\ell)[8_a]}_{gg\to gg} \right) \, ,
\label{elasexpand}
\eeq
with
\beq
M^{(\ell)[8_a]}_{gg\to gg} = \sum_{i=0}^\ell M^{(\ell,i)[8_a]}_{gg\to gg} \log^i\left(\frac{s}{\tau}\right) \,,
\label{elasexpand2}
\eeq
where the $M^{(\ell,\ell)[8_a]}_{gg\to gg}$ coefficients have LL accuracy, the $M^{(\ell,\ell-1)[8_a]}_{gg\to gg}$ coefficients have NLL accuracy,
and in general the $M^{(\ell,\ell-k)[8_a]}_{gg\to gg}$ coefficients have $\mathrm{N^kLL}$ accuracy.

Beyond LL accuracy, in the gluon ladder the exchange of two or more Reggeised gluons may appear.
Further, all the colour representations (\ref{8*8}) exchanged in the $t$ channel may contribute.
However, at NLL accuracy, the real part of the amplitude is entirely given by the antisymmetric octet ${\bf 8}_a$
through \eqn{sudall},
\begin{equation}
 \RE\left[ \cM_{gg\to gg} \right]_{NLL} = \RE\left[ \cM^{(-)[8_a]}_{gg\to gg} \right] \,, 
\end{equation}
which, once \eqn{sudall} is expanded at one and two loops reads,
\beqa
&& \RE\left[ \cM^{(1)}_{gg\to gg} \right]_{NLL} = \alpha^{(1)}(t) \log\left(\frac{s}{\tau}\right) + 2 \RE\big[ C^{g (1)}(t,\tau) \big]\,, \nn\\
&& \RE\left[ \cM^{(2)}_{gg\to gg} \right]_{NLL} \\
&& \qquad\qquad =
\frac1{2} \left( \alpha^{(1)}(t) \right)^2 \log^2\left(\frac{s}{\tau}\right) 
+ \left( \alpha^{(2)}(t) + 2 \RE\big[ C^{g (1)}(t,\tau) \big] \alpha^{(1)}(t) \right) \log\left(\frac{s}{\tau}\right)\,. \nn
\eeqa
Beyond two loops, no more coefficients occur at NLL accuracy, i.e. the gluon-gluon scattering amplitude is uniquely
determined by \eqn{sudall}, in terms of the two-loop Regge trajectory 
$\alpha^{(2)}(t)$~\cite{Fadin:1995xg,Fadin:1996tb,Fadin:1995km,Blumlein:1998ib,DelDuca:2001gu}
and the one-loop impact factor $C^{(1)}$~\cite{Fadin:1993wh,Fadin:1992zt,Fadin:1993qb,DelDuca:1998kx,Bern:1998sc}.
Accordingly, gluon Reggeisation is extended to NLL accuracy~\cite{Fadin:2006bj,Fadin:2015zea}.
Further, because of \eqn{sudall} factorisation still holds, so
the amplitudes for quark-gluon or quark-quark scattering 
have the same form as \eqn{sudall}, up to replacing one or both
colour and impact factors for gluons with the ones for quarks.

\begin{figure}
  \centerline{\includegraphics[width=0.2\columnwidth]{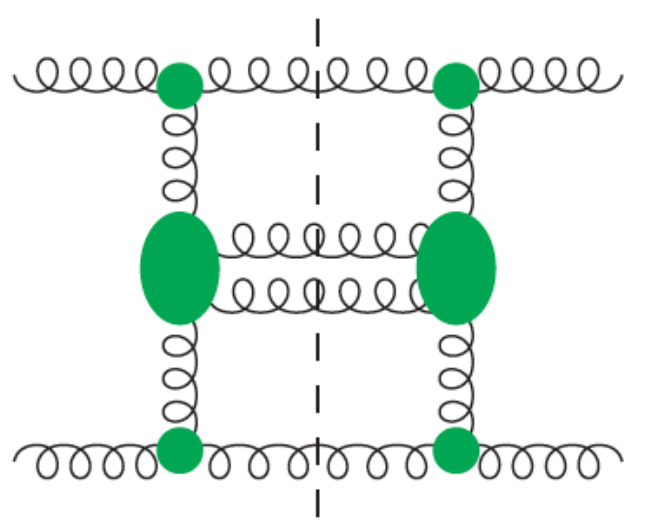} ~~~~
  \includegraphics[width=0.2\columnwidth]{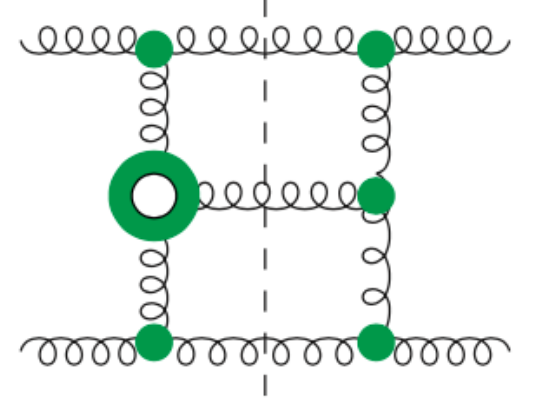} ~~~~
  \includegraphics[width=0.25\columnwidth]{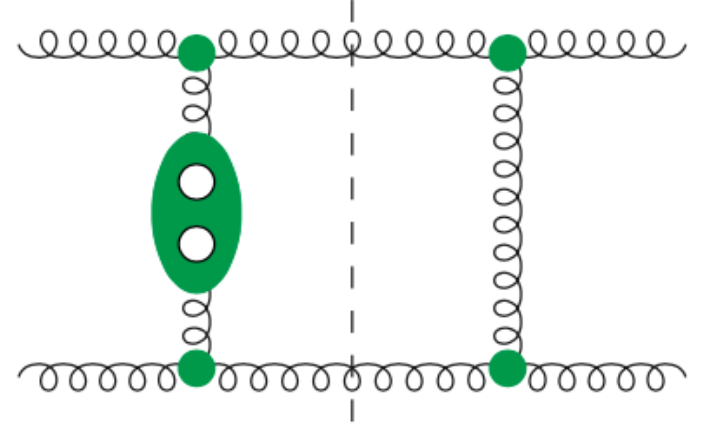} }
  \caption{$(a)$ Tree six--gluon amplitude. The oval blob represents the central-emission vertex --
the emission of two gluons or of a quark-antiquark pair along the gluon ladder.
$(b)$ One-loop five--gluon amplitude. The pierced blob represents the one-loop corrections to 
the central-emission vertex. $(c)$ Regge trajectory in the two-loop four--gluon amplitude. 
The double pierced oval represents the two-loop Regge trajectory.}
\label{fig:lipvert2}
\end{figure}

Likewise, the BFKL equation is extended to NLL accuracy~\cite{Fadin:1998py,Ciafaloni:1998gs},
which sums the terms of ${\cal O}(\as^n \log^{n-1}(s/\tau))$.
At NLL accuracy, the kernel of the BFKL equation is given by
the radiative corrections to the central-emission vertex, i.e.
the emission of two gluons or of a quark-antiquark pair along the 
gluon ladder~\cite{Fadin:1989kf,DelDuca:1995ki,Fadin:1996nw,DelDuca:1996nom,DelDuca:1996km}, 
\fig{fig:lipvert2}$(a)$, and the one-loop corrections to the central-emission 
vertex~\cite{Fadin:1993wh,Fadin:1994fj,Fadin:1996yv,DelDuca:1998cx,Bern:1998sc}, \fig{fig:lipvert2}$(b)$.
The infrared divergences of the radiative corrections to the central-emission vertex
cancel the divergences of the two-loop Regge trajectory, \fig{fig:lipvert2}$(c)$.
The emission of two gluons or of a quark-antiquark pair along the 
gluon ladder requires the adoption of a next-to-multi-Regge kinematics (NMRK),
in which the gluons are strongly ordered in rapidity except for two gluons or for
a quark-antiquark pair emitted along the gluon ladder.
The NMRK rationale is that when the two gluons or the quark-antiquark pair 
are integrated out in rapidity, they yield a factor of ${\cal O}( \as^2 \log(s/\tau) )$, thus
contributing to NLL accuracy.
Finally, in order to compute jet cross sections at NLL accuracy through the BFKL 
equation~\cite{Colferai:2010wu,Ducloue:2013hia,Ducloue:2013bva},
one needs to compute the jet impact factors at next-to-leading order (NLO) 
in $\as$~\cite{Bartels:2001ge,Bartels:2002yj}.
To that effect, one needs the one-loop impact factor $C^{(1)}$ and the impact factor
for the emission of two gluons or of a quark-antiquark 
pair~\cite{Fadin:1989kf,DelDuca:1995ki,Fadin:1996nw,DelDuca:1996nom,DelDuca:1996km}, 
evaluated in NMRK, \sec{sec:appc},
in which the gluons are strongly ordered in rapidity except for two partons emitted
at one end of the ladder.

At NLL accuracy, in addition to the imaginary part of the antisymmetric octet ${\bf 8}_a$ in 
\eqn{sudall}\footnote{In order to avoid an imaginary part in the antisymmetric octet ${\bf 8}_a$,
one can define the combination of logarithms~\cite{Caron-Huot:2017fxr},
\begin{equation}
L = \frac1{2} \left( \log\left(\frac{s}{\tau}\right) + \log\left(\frac{u}{\tau}\right) \right) = \log\left(\frac{s}{\tau}\right) - i\frac{\pi}2\,.
\label{eq:lcross}
\end{equation}
Using $L$ as the resummation parameter,
one may identify odd(even)-signature states with the real (imaginary) part of the amplitude, and thus with the
exchange of an odd (even) number of Reggeised gluons, which is a convenient bookkeping device to classify
the various contributions to the $t$-channel exchange.
At the accuracy of our study, this is immaterial, and we will keep using $\log(s/\tau)$ as the resummation parameter.},
all the colour representations which are even under $s\leftrightarrow u$ crossing (\ref{8*8}),
i.e. the singlet ${\bf 1}$, the symmetric octet ${\bf 8}_s$ and the ${\bf 27}$,
appear in the imaginary part of the gluon-gluon scattering amplitude,
already in the non-logarithmic term of the one-loop amplitude $M^{(1,0)}_{gg\to gg}$~\cite{DelDuca:1998kx}.
Just like the BFKL equation~\cite{Kuraev:1976ge,Kuraev:1977fs,Balitsky:1978ic}, which
is described by the exchange of two Reggeised gluons in the $t$ channel and displays a cut in the $s$ channel,
the imaginary part of the gluon-gluon amplitude features a cut
and is described by the exchange of two Reggeised gluons~\cite{Caron-Huot:2013fea,Caron-Huot:2017zfo}, and just like the
BFKL equation~\cite{DelDuca:2013lma} its finite parts can be expressed in terms of single-valued
harmonic polylogarithms~\cite{Caron-Huot:2020grv}.
The analyses of Refs.~\cite{Caron-Huot:2017zfo,Caron-Huot:2020grv} complete the study of
gluon-gluon scattering amplitudes at NLL accuracy.
We note that the imaginary part of the gluon-gluon amplitude at NLL accuracy can only contribute
to a squared amplitude, and thus to a cross section, at next-to-next-to-leading-logarithmic (NNLL) accuracy.

\subsection{Moving toward next-to-next-to-leading logarithmic accuracy}
\label{sec:reggennll}

At NNLL accuracy, both the antisymmetric octet ${\bf 8}_a$ (starting at two loops) and the 
${\bf 10} \oplus \overline{{\bf 10}}$ (starting at three loops) feature a cut in the $s$ channel,
which is described by the exchange of three Reggeised gluons in the $t$ channel~\cite{Fadin:2016wso,Caron-Huot:2017fxr,Fadin:2017nka}.
This is in addition to the exchange of one Reggeised gluon, which is still given by \eqn{sudall}.
Thus, Regge pole factorisation, which is based on the exchange of one Reggeised gluon, does not hold any more~\cite{DelDuca:2001gu}. 
Accordingly, in the non-logarithmic term of the two-loop amplitude, $M^{(2,0)}_{gg\to gg}$, both the two-loop impact factor
and the three-Reggeised-gluon 
exchange\footnote{At NNLL accuracy, the three-Reggeised-gluon exchange is currently known up to four loops~\cite{Falcioni:2020lvv}.} 
contribute.

However, in $M^{(2,0)}_{gg\to gg}$, the factorisation-violating term is a $1/\eps^2$ pole, 
whose coefficient is subleading in $N_c$~\cite{DelDuca:2001gu}.
Since it is an infrared pole, it can be analysed through infrared factorisation, which shows that,
in addition to the usual diagonal terms of the colour octet exchange, also
non-diagonal terms in the $t$-channel colour basis contribute to the real part of $M^{(2,0)}_{gg\to gg}$~\cite{Bret:2011xm,DelDuca:2011ae}. 
Then one can disentangle the infrared poles of the two-loop impact factor (\ref{sudall})
from the ones of factorisation-violating terms by identifying the impact factor with the diagonal terms and the factorisation-violating terms
with the off-diagonal ones. 
Carrying this prescription on to the next loop order,
one can predict how the infrared poles of the factorisation-violating term propagate into the
single-logarithmic term of the three-loop amplitude $M^{(3,1)}_{gg\to gg}$~\cite{DelDuca:2013ara,DelDuca:2014cya},
and thus have a prescription to disentangle the infrared poles of the three-loop Regge trajectory,
which should be given by the cusp anomalous dimension~\cite{Korchemskaya:1996je}, from the ones of the factorisation-violating terms.
The disentanglement, which is based on such a prescription, has been 
confirmed by the explicit computation of the three-loop four-point function of ${\cal N}=4$ SYM~\cite{Henn:2016jdu}.
We stress, though, that the disentanglement above is valid only for the infrared poles through three loops.
In order to predict factorisation-violating terms beyond three loops and including the finite terms, 
one must be able to compute the three Reggeised-gluon exchange~\cite{Caron-Huot:2017fxr,Caron-Huot:2020grv}.

The possibility of disentangling terms which are based on the exchange of one Reggeised gluon
from factorisation-violating terms makes us hope that the BFKL equation,
which is based on the exchange of one Reggeised gluon, could be extended to NNLL accuracy.
Further, there are reasons, which are based on the integrability of amplitudes in MRK in the large $N_c$ limit~\cite{Lipatov:2009nt}
and on the colour structure of the corrections found through four loops in Ref.~\cite{Falcioni:2020lvv}, to believe
that the Regge pole factorisation would be simpler in the large $N_c$ limit.
This warrants an analysis of the terms which would contribute to the BFKL equation at NNLL accuracy.

At NNLL accuracy, the kernel of the BFKL equation will be given by
the emission of three partons along the gluon ladder~\cite{DelDuca:1999iql,Antonov:2004hh,Duhr:2009uxa},
evaluated in next-to-next-to-multi-Regge kinematics (NNMRK);
by the one-loop corrections to the emission of two gluons or 
of a quark-antiquark pair along the gluon ladder, evaluated in NMRK;
and by the two-loop corrections to the central-emission vertex.
The last two contributions are yet to be determined.
The infrared divergences of the contributions outlined above will
cancel the divergences of the three-loop Regge trajectory, 
which is so far only known in a specific scheme~\cite{Caron-Huot:2017fxr}.

Finally, in order to compute jet cross sections at NNLL accuracy through the BFKL 
equation, one needs to compute the jet impact factors at next-to-next-to-leading order (NNLO) 
in $\as$.
To that effect, one needs the tree impact factor for the emission of three partons at one end of the ladder~\cite{DelDuca:1999iql},
evaluated in NNMRK; the two-loop impact factor $C^{(2)}$~\cite{DelDuca:2014cya},
and the one-loop impact factor for the emission of two gluons or of a quark-antiquark 
pair, evaluated in NMRK. Further, one must include the square of the one-loop helicity-violating impact factor, which is recalled 
in \eqns{eq:helviol}{eq:1lhelviol}.

\subsection{Plan}

Using the one-loop corrections to the five-gluon amplitudes~\cite{Bern:1993mq} 
and to the two-quark three-gluon amplitudes~\cite{Bern:1994fz},
in this paper we evaluate the one-loop impact factor for the emission of two gluons.
This is the last ingredient which is necessary to evaluate the gluon-jet impact factor at NNLO accuracy in $\as$.

Sec.~\ref{sec:if} is devoted to the computation of the one-loop impact factor for the emission of two gluons.
Firstly, we recall the tree-level impact factor; then in \sec{sec:5g1lamp}, we consider the one-loop five-gluon amplitude in NMRK, 
and apply the reggeisation ansatz to each colour-ordered amplitude (\ref{NLLfactorization});
from that we extract in \sec{sec:1loopif2g}
the one-loop impact factor for the emission of two gluons (\ref{oneloopIFgg}). These are the main results of the paper.
Then, in order to verify that Regge factorisation holds at NLL accuracy in NMRK, we consider in \sec{sec:reggefact}
the one-loop two-quark three-gluon amplitude in NMRK,
and we extract again the one-loop impact factor for the emission of two gluons, finding agreement with the previous computation. Finally,
in \sec{sec:helviol1loopif} we compute the one-loop helicity-violating impact factor.
In \sec{sec:conc}, we draw our conclusions. Several appendices are given, which contain information about the kinematic approximations,
and the one-loop five-point amplitudes in general kinematics and in NMRK.

\section{The impact factor for the emission of two gluons}
\label{sec:if}

\begin{figure}
  \centerline{\includegraphics[width=0.5\columnwidth]{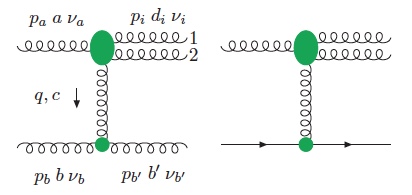} }
  \caption{Tree $(a)$ five--gluon amplitude for the $g_a\, g_b\to g_1\, g_2\,g_{b'}$ scattering,
  and $(b)$ two-quark three--gluon amplitude for the $g_a\, q_b\to g_1\, g_2\,q_{b'}$ scattering, in the NMRK of \eqn{qmrapp1}. 
  The oval blob represents the impact factor for the emission of two gluons.}
\label{fig:ifggtree}
\end{figure}

We consider the production of three gluons of momenta $p_1$, $p_2$ and $p_{b'}$ in the scattering 
of two gluons of momenta $p_a$ and $p_b$. We suppose that the outgoing gluons are in NMRK, \sec{sec:appc},
with rapidities and transverse momenta,
\begin{equation}
y_1 \simeq y_2 \gg y_{b'}\,;\qquad |p_{1\perp}|
\simeq |p_{2\perp}| \simeq |p_{b'\perp}|\, .
\label{qmrapp1}
\end{equation}
In the NMRK region of \eqn{qmrapp1}, the tree amplitude for the $g_a\, g_b\to g_1\, g_2\,g_{b'}$ scattering is~\cite{DelDuca:1996km}
%
%
\beq
 \left. \cM^{(0) }_{gg\to ggg} \right|_{NMRK} = \sum_{\sigma \in S_2}\gs^3\, (F^{d_{\sigma_1}}F^{d_{\sigma_2}})_{ac} (F^{b'})_{cb} \,
 \left. m^{(0)}_{5g}(p_a^{\nu_a}, p^{\nu_{\sigma_1}}_{\sigma_1}, p^{\nu_{\sigma_2}}_{\sigma_2}, p_{b'}^{\nu_{b'}}, p_b^{\nu_b})\right|_{NMRK} \,,
 \label{NLOfactorization}
\eeq
where the sum is over the permutations of the labels 1 and 2, and the colour-ordered amplitude is
\beq
 \left. m^{(0)}_{5g}(p_a^{\nu_a}, p^{\nu_1}_1, p^{\nu_2}_2, p_{b'}^{\nu_{b'}}, p_b^{\nu_b})\right|_{NMRK} =
A^{gg(0)}(p_a^{\nu_a}; p^{\nu_1}_1, p^{\nu_2}_2)\, \frac{s}{t}\,  C^{g(0)}(p_b^{\nu_b};p_{b'}^{\nu_{b'}}) \,, 
\label{NLOfactorization2}
\eeq
with $s=p_a+p_b$, $q = p_{b'} + p_b$ and $t \simeq -|q_\perp|^2$, \fig{fig:ifggtree}(a). 
\Eqn{NLOfactorization} yields the five-gluon amplitude in terms of two colour-ordered amplitudes, again in agreement with the Bern-Carrasco-Johansson (BCJ) counting of $(n-3)!$ independent colour-ordered amplitudes.

In \eqn{NLOfactorization2}, the impact factor $A^{gg(0)}(p_a^{\nu_a}; p^{\nu_1}_1, p^{\nu_2}_2 )$ for $g^*\, g \rightarrow g\, g$ can be written as
\begin{equation}
A^{gg(0)}(p_a^{\nu_a}; p^{\nu_1}_1, p^{\nu_2}_2 ) =
D^{gg}(p_a^{\nu_a}; p^{\nu_1}_1, p^{\nu_2}_2 ) B^{\tilde\nu}(p_1,p_2) \,,
\label{eq:acoeff}
\end{equation}
with
\beqa
D^{gg}(p_a^-; p^+_1, p^+_2 ) &=& 1\,, \nn\\
D^{gg}(p_a^+; p^-_1, p^+_2 ) &=& x_1^2\,, \label{eq:dcoeff}\\
D^{gg}(p_a^+; p^+_1, p^-_2 ) &=& x_2^2\,, \nn
\eeqa
and 
\beq
\tilde\nu = {\rm sign}(\nu_a + \nu_1 + \nu_2)\,,
\label{eq:nutilde}
\eeq
and\footnote{Note that since $(F^c)_{ab} = i\sqrt{2} f^{acb}$, the coefficient (\ref{eq:fcoeff}) and the central-emission vertex (\ref{eq:lipv})
do not display the usual overall $\sqrt{2}$ factor. }
\beq
B^+(p_1,p_2) = \frac{p_{1\perp}+p_{2\perp}}{p_{1\perp}} \sqrt{\frac{x_1}{x_2}} \frac1{\langle 12\rangle} \,,
\label{eq:fcoeff}
\eeq
and where the momentum fractions are
\beq
x_i = \frac{p_i^+}{p_1^+ + p_2^+}\,, \qquad i=1, 2\,.
\label{sec:x12}
\eeq
The $B$ coefficient (\ref{eq:fcoeff}) displays a collinear divergence as $p_1\cdot p_2\to 0$. 
The $D$ coefficients (\ref{eq:dcoeff}) are real, and so invariant under parity, while $B$  transforms under parity into its
complex conjugate,
\beq
B^-(p_1,p_2) = [B^+(p_1,p_2)]^\ast\,.
\eeq
The impact factor $A^{gg(0)}(p_a^{\nu_a}; p^{\nu_2}_2 , p^{\nu_1}_1)$ is obtained from \eqn{eq:acoeff} by swapping the labels 1 and 2.

Since the five-gluon amplitude is a MHV amplitude, \eqn{NLOfactorization} describes $2\binom{5}{2} = 20$ 
helicity configurations. However, at leading power in $t/s$, only 12 helicity configurations are leading. 
The impact factor (\ref{eq:acoeff}) describes six helicity configurations, three out of which are given by \eqn{eq:dcoeff}, and three more by
parity conjugation. They are multiplied by the two given by the impact factor $C^{g(0)}(p_b^\mp;p_{b'}^\pm)$ (\ref{centrc}).
Then there are eight helicity configurations which are power suppressed in $t/s$.
Three of them are associated to $C^{g(0)}(p_b^-;p_{b'}^-)$, multiplied by the impact factor (\ref{eq:acoeff})
with the helicity configurations (\ref{eq:dcoeff}), one is associated to $C^{g(0)}(p_b^-;p_{b'}^-)$ multiplied by
$A^{gg(0)}(p_a^+; p^+_1, p^+_2 )$. The remaining four are obtained from the four above by parity conjugation.

In MRK, \app{sec:appb}, with $p_1^+\gg p_2^+$, the impact factor $A^{gg(0)}(p_a^{\nu_a}; p^{\nu_2}_2, p^{\nu_1}_1 )$ 
is power suppressed in $p_2^+/p_1^+$ because of the $F$ coefficient (\ref{eq:fcoeff}), while
$A^{gg(0)}(p_a^{\nu_a}; p^{\nu_1}_1, p^{\nu_2}_2 )$ factorises as
\beq
\lim_{ p_1^+\gg p_2^+ } A^{gg(0)}(p_a^{\nu_a}; p^{\nu_1}_1, p^{\nu_2}_2 ) = 
C^{g(0)}(p_a^{\nu_a}; p^{\nu_1}_1)\, \frac1{t_1}\, V^{g(0)}(q_1,p^{\nu_2}_2)\,,
\eeq
where $q_1 = - (p_a+p_1)$, $q_2= q_1-p_2$, $t_1= q_1^2\simeq - q_{1\perp} q_{1\perp}^\ast$,
and where $C^{g(0)}(p_a^{\nu_a}; p^{\nu_1}_1)$ is given in \eqn{centrc}, while the tree central-emission vertex 
is
\beq
V^{g(0)}(q_1,p^+_2) = \frac{q_{1\perp}^\ast q_{2\perp}}{p_{2\perp}}\,.
\label{eq:lipv}
\eeq
Accordingly, the five-gluon amplitude (\ref{NLOfactorization}) takes the factorised form,
\begin{eqnarray}
\left. \cM^{(0) }_{gg\to ggg} \right|_{MRK} &=& 
s \left[\gs (F^{d_1})_{ac_1}\, C^{g(0)}(p_a^{\nu_a};p_1^{\nu_1}) \right]\, 
{1\over t_1}\, \label{three}\\ &\times& \left[\gs (F^{d_2})_{c_1c_2}\, 
V^{g(0)}(q_1,p^{\nu_2}_2) \right]\, {1\over t_2}\, 
\left[\gs (F^{b'})_{c_2b}\, C^{g(0)}(p_b^{\nu_b};p_{b'}^{\nu_{b'}}) \right]\, .\nonumber
\end{eqnarray}

\subsection{The one-loop five-gluon amplitude in the next-to-multi-Regge kinematics}
\label{sec:5g1lamp}

In order to extend \eqn{NLOfactorization} to loop level and to NLL accuracy, we note that
the five-gluon amplitude in the NMRK region of \eqn{qmrapp1} retains the crossing symmetry under $p_b \leftrightarrow p_{b'}$.
Using \eqn{nmrinv}, that is equivalent to a crossing symmetry of the colour-ordered amplitudes of \eqn{NLOfactorization}
under either $s_{2b} \leftrightarrow s_{2b'}$ or $s_{1b} \leftrightarrow s_{1b'}$.
%
%
Then, just like for the four-gluon
amplitude (\ref{eq:sucross}), but in this case for each colour-ordered amplitude,
we may consider states whose kinematic and colour coefficients have a definite signature 
under the suitable crossing, extend the reggeisation ansatz (\ref{sudall}) to NMRK, and write
%
\beq
\left. \cM^{(-)[8_a]}_{gg\to ggg}\right|_{NMRK}
= \sum_{\sigma \in S_2}\gs^3\, (F^{d_{\sigma_1}}F^{d_{\sigma_2}})_{ac} (F^{b'})_{cb} \,
 \left. m_{5g}(p_a^{\nu_a}, p^{\nu_{\sigma_1}}_{\sigma_1}, p^{\nu_{\sigma_2}}_{\sigma_2}, p_{b'}^{\nu_{b'}}, p_b^{\nu_b})\right|_{NMRK} \,,
\label{NLLfactorization}
\eeq
where, as in \eqn{NLOfactorization}, the sum is over the permutations of the labels 1 and 2, and
\beqa
\lefteqn{ \left. m_{5g}(p_a^{\nu_a}, p^{\nu_1}_1, p^{\nu_2}_2, p_{b'}^{\nu_{b'}}, p_b^{\nu_b})\right|_{NMRK} } \nn\\
&=& \frac1{2} {s\over t} A^{gg}(p_a^{\nu_a}; p^{\nu_1}_1, p^{\nu_2}_2) 
\left[ \left({s_{2 b'}\over \tau}\right)^{\alpha(t)} + \left({-s_{2 b'}\over \tau}\right)^{\alpha(t)} \right] 
C^{g}(p_b^{\nu_b};p_{b'}^{\nu_{b'}}) \,,  
\label{NLLfactorizatio2}
\eeqa
where $\alpha(t)$ is given by \eqn{alphb}, $C^{g}(p_b^{\nu_b};p_{b'}^{\nu_{b'}})$ by \eqn{fullv}, and
\beq
A^{gg}(p_a^{\nu_a}; p^{\nu_1}_1, p^{\nu_2}_2 ) = 
A^{gg(0)}(p_a^{\nu_a}; p^{\nu_1}_1, p^{\nu_2}_2 ) \left(1 + \frac{\as}{4\pi} A^{gg(1)}_{\tilde\nu}(p_1,p_2,\tau) + \ord(\as^2) \right)\, ,
\label{eq:a1loop}
\eeq
with $\tilde\nu$ in \eqn{eq:nutilde}. Just like in \eqn{elasexpand}, one can write the colour-ordered amplitude (\ref{NLLfactorizatio2}) 
as a double expansion in the strong coupling $\as$ and in $\log(s_{2 b'}/\tau)$, 
\beq
\left. m_{5g}\right|_{NMRK} =  m^{(0)}_{5g}
\left( 1 + \sum_{\ell=1}^\infty \left(\frac{\as}{4\pi}\right)^\ell m^{(\ell)}_{5g} \right) \, ,
\label{nllnmrkexpand}
\eeq
with $m^{(0)}_{5g}$ as in \eqn{NLOfactorization2} where a $NMRK$ subscript is understood,
and where
\beq
m^{(\ell)}_{5g} = \sum_{i=0}^\ell \log^i\left(\frac{s_{2 b'}}{\tau}\right) m^{(\ell,i)}_{5g}   \,,
\label{nllnmrkexpand2}
\eeq
where the $m^{(\ell,\ell)}_{5g}$ coefficients have LL accuracy, the $m^{(\ell,\ell-1)}_{5g}$ 
coefficients have NLL accuracy, etc.

\begin{figure}
  \centerline{\includegraphics[width=0.7\columnwidth]{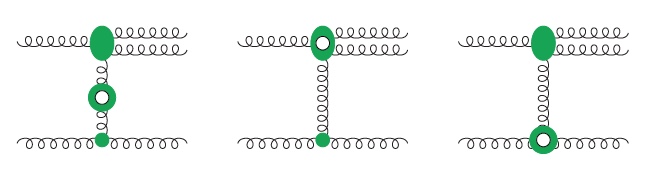} }
  \caption{Factorisation of the one-loop five-gluon amplitude in NMRK.
  The three diagrams represent the three terms in \eqn{eq:1loopsubtr}.
  The pierced blobs represent $(a)$ the one-loop Regge trajectory;
  and the one-loop corrections to the impact factor for the emission of $(b)$ two gluons and $(c)$ one gluon.}
\label{fig:ifgg1loop}
\end{figure}

If at NLL accuracy only the antisymmetric octet ${\bf 8}_a$ is exchanged  in the $t$ channel,
and if within ${\bf 8}_a$ three reggeised-gluon exchanges are missing,
we expect that, just like for \eqn{sudall}, the real part of the five-gluon amplitude in the NMRK region of \eqn{qmrapp1}
be entirely given by \eqn{NLLfactorization},
\begin{equation}
 \RE\left[ \cM_{gg\to ggg} \right]^{NLL}_{NMRK} = \RE\left[ \cM^{(-)[8_a]}_{gg\to ggg} \right]_{NMRK}\,,
 \label{eq:nllnmrkm5}
\end{equation}
where the left-hand side of \eqn{eq:nllnmrkm5}, once expanded at one loop, yields
\beq
\RE\left[ \cM^{(1) }_{gg\to ggg} \right]_{NMRK}^{NLL} 
= \gs^3\, \frac{\as}{4\pi}\, \sum_{\sigma \in S_2} (F^{d_{\sigma_1}}F^{d_{\sigma_2}} F^{b'})_{ab} \,
\RE\left[ m_{5:1} \right]_{NMRK}\,,
\label{eq:nllnmrkm5lhs}
\eeq
where the colour-ordered amplitude $m_{5:1}$ will be provided in \app{sec:1loopcolamp} and \app{sec:1loopampnmrk},
where the dependence on the momenta to be permuted is understood,
and where, using \eqns{nllnmrkexpand}{eq:nllnmrkm5}, we write
\beq
\RE\left[ m_{5:1}\right]_{NMRK} =  m^{(0)}_{5g}\, \RE\left[ m^{(1)}_{5g} \right]_{NMRK} \,,
\label{eq:1loopsubtr}
\eeq
with
\beq
\RE\left[ m^{(1)}_{5g} \right]_{NMRK} =
\alpha^{(1)}(t) \log\left(\frac{s_{2 b'}}{\tau}\right) + \RE\big[ A^{gg(1)}(p_1, p_2,\tau) \big] 
+ \RE\big[ C^{g (1)}(t,\tau) \big] \,,
\label{eq:1loopsubtr2}
\eeq
which is displayed diagrammatically in \fig{fig:ifgg1loop}.
The one-loop impact factors $C^{p (1)}$, with $p = g, q$, are process and infrared-scheme dependent.
They were computed in conventional dimensional 
regularization (CDR)/'t-Hooft-Veltman (HV) schemes in 
Refs.~\cite{Fadin:1992zt,Fadin:1993wh,Fadin:1993qb,DelDuca:1998kx,Bern:1998sc}, and
in the dimensional reduction (DR)/ four dimensional helicity (FDH) schemes in Refs.~\cite{Bern:1998sc,DelDuca:1998kx}.
Through $\ord(\eps^0)$, the real part of the $\overline{\rm MS}$-renormalised one-loop gluon impact factor can be written 
as~\cite{DelDuca:1998kx,DelDuca:2017pmn}
\beqa
\RE\big[ C^{g (1)}_{-+}(t,\tau) \big] &=&  \cg \left\{ \left({\mu^2\over -t}\right)^{\eps} \left[
- \frac{\gamma_K^{(1)} }{\epsilon^2} N_c + \frac{4\gamma_g^{(1)} }{\epsilon} + \frac{\beta_0}{2\eps} 
+ \frac{N_c}{\eps} \log\left(\frac{\tau}{-t}\right)  \right.\right. \nn\\
&& \left.\left. \qquad\qquad\quad
- \left( \frac{32}{9} + \frac{\delta_R}{6} - \frac{\pi^2}{2} \right) N_c + \frac{5}{9} N_f \right] -  \frac{\beta_0}{2\eps} \right\} \nn\\
&=&  \cg  \left[ \left( \frac{\mu^2}{-t} \right) ^\epsilon 
\left[ \left( -\frac{2}{\epsilon^2} -\frac{11}{6\eps} + \frac1{\eps} \log\left(\frac{\tau}{-t}\right)
- \frac{32}{9} - \frac{\delta_R}{6} + \frac{\pi^2}{2} \right) N_c \right.\right. \nn\\
&& \left.\left. \qquad\qquad\quad
+ \left( \frac1{3\eps} + \frac{5}{9} \right) N_f \right] - \frac{\beta_0}{2\epsilon} \right] \,,
\label{impactcorrect}
\eeqa
with $N_f$ the number of light quark flavours, and where
we have used the regularisation parameter, $\delta_R =1$ in CDR/HV schemes, $\delta_R =0$ in the DR/FDH  schemes~\cite{Catani:1996pk}.
In \eqn{impactcorrect}, the last term is the $\overline{\rm MS}$ ultraviolet counterterm,
and the infrared $\eps$ poles are accounted for by the cusp anomalous dimension
and by the gluon collinear anomalous dimension~\cite{DelDuca:2014cya},
with $\gamma_K^{(1)}$ the one-loop coefficient of the cusp anomalous dimension (\ref{eq:k2}), 
$\gamma_g^{(1)}$ the one-loop coefficient of the 
gluon collinear anomalous dimension (\ref{eq:c1}), and $\beta_0$ the one-loop coefficient of the beta function (\ref{eq:b0k2}).

In order to compute $A^{gg(1)}$, we use the colour decomposition of the one-loop five-gluon amplitude~\cite{DelDuca:1999rs},
\beqa
\cM^{(1)}_{5g} &=& \gs^{5} \sum_{\sigma \in S_4/\cR} 
\left[ \Tr (F^{d_{\sigma_1}} \cdots F^{d_{\sigma_5}}) m_{5:1}^{[1]}(\sigma_1, \ldots, \sigma_5) \right. \nn\\
&& \left. \qquad + 2N_f \Tr (T^{d_{\sigma_1}} \cdots T^{d_{\sigma_5}}) m_{5:1}^{[1/2]}(\sigma_1, \ldots, \sigma_5) \right] \,,
\label{eq:oneLcol}
\eeqa
where $S_4 \equiv S_{5}/\mathbb{Z}_{5}$ is the group of the non-cyclic permutations, and $\cR$ is the reflection:
$\cR(1,\ldots,5) = (5,\ldots,1)$. The superscript $[j]$ denotes the spin of the particle circulating in the loop.
\Eqn{eq:oneLcol} contains 12 independent colour-ordered amplitudes of type $m_{5:1}^{[1]}$
and 12 of type $m_{5:1}^{[1/2]}$. They have been computed in Ref.~\cite{Bern:1993mq}, and are recorded in \app{sec:1loopcolamp}
in the notation of Ref.~\cite{DelDuca:1998cx}.

We evaluate the colour-ordered amplitudes $m_{5:1}^{[1]}$ and $m_{5:1}^{[1/2]}$, \app{sec:1loopcolamp}, 
in the NMRK region of \eqn{qmrapp1} and \app{sec:appc}. The leading colour-ordered amplitudes are found to be the same as at tree 
level~(\ref{NLOfactorization}), such that at one loop we can write the five-gluon colour-ordered amplitude as in \eqn{eq:nllnmrkm5lhs},
where $m_{5:1}$ is given by \eqns{eq:colweight}{oneloop1}.
Then the one-loop five-gluon amplitude is factorised as in \eqnss{NLLfactorization}{eq:1loopsubtr}.

\subsection{One-loop impact factor for the emission of two gluons}
\label{sec:1loopif2g}

Using eqs.~(\ref{alph}), (\ref{eq:1loopsubtr}) and (\ref{impactcorrect}), we obtain from \eqn{oneloop1} the real part of the 
$\overline{\rm MS}$-renormalised one-loop impact factor for the emission of two gluons,
\beqa
\lefteqn{ \frac1{\cg} \RE\big[ A^{gg(1)}_+(p_1,p_2,\tau) \big]  }\nn\\
&=& N_c \left\{ -\frac{1}{\epsilon^2} \left[ 2 \left( \frac{\mu^2}{|p_{1\perp}|^2} \right)^\epsilon 
+ \left( \frac{\mu^2}{|p_{2\perp}|^2} \right)^\epsilon \right] \right. \nn\\ 
&&\qquad + \frac1{\epsilon} \left[ 2 \left(\frac{\mu^2}{|p_{1\perp}|^2}\right)^\epsilon \log\left(\frac{s_{12}}{|p_{1\perp}| \: |p_{2\perp}|}\right)
+ \left(\frac{\mu^2}{-t}\right)^\epsilon \log\left(\frac{\tau }{|p_{2\perp}|^2}\right) \right] \nn\\
&& \qquad - \frac1{\epsilon} \left( \frac{\mu^2}{-t}\frac{|p_{2\perp}|^2}{|p_{1\perp}|^2} \left( \frac{p_2^+}{p_1^+}\frac{s_{12}}{|p_{2\perp}|^2} \right)^{\frac{1}{2}} \right)^\epsilon  \log \left( \frac{x_1x_2s_{12}}{|p_{2\perp}|^2} \right) \nn\\
&& \left. \qquad   -\frac{1}{2} \log^2\left(\frac{|p_{1\perp}|^2}{-t}\right) 
- \frac{1}{2}\log\left(x_1 \right)\log\left(x_1x_2\left(\frac{-t}{s_{12}}\right)^3 \frac{-t}{|p_{2\perp}|^2} \right) 
- \frac{32}{9} - \frac{\delta_R}{6} + \frac{5}{6}\pi^2\right\} \nn\\
&-& \frac{\beta_0}{2\epsilon} \left(\frac{x_1 \mu^2}{|p_{1\perp}|^2 }\right)^\epsilon 
+ \frac{5}{9}N_f + \frac{\beta_0}{2}\left(\frac{|p_{1\perp}|^2}{x_1} - t + 2q_\perp^*p_{1\perp}\right) \frac{L_0\left(\frac{|p_{1\perp}|^2}{-x_1t}\right)}{t} \nn\\
&+& \frac{N_C-N_f}{3} \left\{ \frac{p_2^+}{p_1^+}\left[ |p_{1\perp}|^2 \: {q_\perp^*}^2 \: p_{2\perp}^2
-2\: |p_{1\perp}|^2 \: |p_{2\perp}|^2 p_{1\perp} q_\perp^*  + \frac{p_2^+}{p_1^+} |p_{1\perp}|^4 \, p_{2\perp} q_\perp^*\right] \right. \nn\\
&& \qquad\qquad  - |p_{2\perp}|^2  \left[2 \: |p_{1\perp}|^2(-t)+\left( 2\,|p_{2\perp}|^2 - |p_{1\perp}|^2 + t\right) p_{1\perp} q_\perp^*\right] \bigg\}
\frac{L_2\left(\frac{|p_{1\perp}|^2}{-x_1t}\right)}{t^3} \nn\\
&+& \frac{N_C-N_f}{6}\left[\frac{x_2 p_{2\perp} q_\perp^*}{ -t} + \frac{2x_1x_2 {q_\perp^*}^2 p_{1\perp}^2}{- t\, |p_{1\perp}|^2} 
- \frac{ x_1 |p_{2\perp}|^2 p_{1\perp} q_\perp^*}{ -t\, |p_{1\perp}|^2} \right] 
-  \frac{\beta_0}{\epsilon} + O(\epsilon) \,,
\label{oneloopIFgg}
\eeqa
with $q_\perp =  - (p_{1\perp} + p_{2\perp})$, where the last term is the $\overline{\rm MS}$ ultraviolet counterterm,
and where it is understood that in order to obtain the real part we must use \eqn{eq:realpart}.
The impact factor $A^{gg(1)}(p_a^{\nu_a}; p^{\nu_2}_2, p^{\nu_1}_1 )$ for the second colour ordering of \eqn{NLLfactorization}
is obtained by exchanging the labels 1 and 2 in \eqn{oneloopIFgg}.

In MRK, \app{sec:appb}, with $p_1^+\gg p_2^+$, the impact factor $A^{gg}(p_a^{\nu_a}; p^{\nu_2}_2, p^{\nu_1}_1 )$ 
is power suppressed in $p_2^+/p_1^+$, then we just need to consider the first colour ordering in \eqn{NLLfactorization},
for which the one-loop impact factor (\ref{oneloopIFgg}) factorises as
\begin{equation}
\lim_{ p_1^+\gg p_2^+ } \RE\big[ A^{gg(1)}(p_1,p_2,\tau) \big] = \alpha^{(1)}(t_1) \log\left(\frac{s_{12}}{\tau}\right)
+ \RE\big[ C^{g (1)}(t_1,\tau) \big] + \RE\big[ V^{g(1)}(q_1,p_2,\tau) \big] \,,
\label{eq:mrklipvert}
\end{equation}
with $q_1 = - (p_a+p_1)$, $t_1= q_1^2\simeq - q_{1\perp} q_{1\perp}^\ast$ and $q = q_2= q_1-p_2$, and
where the one-loop corrections to the $\overline{\rm MS}$-renormalised central-emission vertex (\ref{eq:lipv}) are given by
\beq
V^{g}(q_1,p_2) = V^{g(0)}(q_1,p_2) \left(1 + \frac{\as}{4\pi} V^{g(1)}(q_1,p_2,\tau) + \ord(\as^2) \right)\, ,
\eeq
with~\cite{DelDuca:1998cx}
\begin{eqnarray}
\frac1{\cg} \lefteqn{ \RE\big[ V^{g(1)}(q_1,p_2,\tau) \big] } \nn\\
&=& N_c \left\{ - {1\over\epsilon^2}\, \left({\mu^2\over |p_{2\perp}|^2}\right)^\epsilon
+ {1\over\epsilon}\, \left[ \left({\mu^2\over -t_1}\right)^\epsilon + \left({\mu^2\over -t_2}\right)^\epsilon \right]
\log\left( {\tau\over |p_{2\perp}|^2} \right)
- {1\over 2} \log^2\left( {t_1\over t_2} \right) + {\pi^2\over 3} \right\} \nn\\
&-& {\beta_0\over 2} \left(t_1 + t_2 + 2 q_{1\perp}
q_{2\perp}^\ast\right) {L_0(t_1/t_2)\over t_2} \nn\\ 
&+& {N_c-N_f\over 3}\, |p_{2\perp}|^2
\left[ - [ 2t_1 t_2 + (t_1 + t_2 + 2|p_{2\perp}|^2) q_{1\perp} 
q_{2\perp}^\ast ] {L_2(t_1/t_2)\over t_2^3} - {q_{1\perp} 
q_{2\perp}^\ast \over 2\, t_1\, t_2} \right] - {\beta_0\over 2\epsilon} \nn\\
 &+& O(\epsilon)\, ,\label{looplip}
\end{eqnarray}
with $t = t_2 = q_2^2$, where the last term is the $\overline{\rm MS}$ ultraviolet counterterm, 
and where it is understood that in order to obtain the real part we must use
\beq
2 {\rm Re}(q_{1\perp}q_{2\perp}^\ast)= - (t_1+t_2+|p_{2\perp}|^2) \,.
\eeq

\subsection{One-loop two-quark three-gluon amplitude in NMRK}
\label{sec:reggefact}

Regge factorisation at tree level implies that in the NMRK region of \eqn{qmrapp1}, 
we can write the two-quark three-gluon tree amplitude for the $g_a\, q_b\to g_1\, g_2\,q_{b'}$ scattering, \fig{fig:ifggtree}(b),
as~\cite{DelDuca:1996km,DelDuca:1999iql}
\beq
 \left. \cM^{(0) }_{gq\to ggq} \right|_{NMRK} = \sum_{\sigma \in S_2}\gs^3\, (F^{d_{\sigma_1}}F^{d_{\sigma_2}})_{ac} (T^c)_{b'\bar{b}} \,
 \left. m^{(0)}_{2q3g}(p_a^{\nu_a}, p^{\nu_{\sigma_1}}_{\sigma_1}, p^{\nu_{\sigma_2}}_{\sigma_2}, p_{b'}^{\nu_{b'}}, p_b^{\nu_b})\right|_{NMRK} \,,
 \label{NLOfactorizationqq}
\eeq
where the sum is over the permutations of the labels 1 and 2, and the colour-ordered amplitude is
\beq
 \left. m^{(0)}_{2q3g}(p_a^{\nu_a}, p^{\nu_1}_1, p^{\nu_2}_2, p_{b'}^{\nu_{b'}}, p_b^{\nu_b})\right|_{NMRK} =
A^{gg(0)}(p_a^{\nu_a}; p^{\nu_1}_1, p^{\nu_2}_2)\, \frac{s}{t}\,  C^{q(0)}(p_b^{\nu_b};p_{b'}^{\nu_{b'}}) \,, 
\label{NLOfactorizationqq2}
\eeq
where $A^{gg(0)}$ is given by \eqn{eq:acoeff}, and 
the quark impact factor is\footnote{Since helicities are for all outgoing momenta, an incoming right-handed quark is labelled in
\eqn{centrcqq} and the right-hand side of \eqn{NLOfactorizationqq} as a left-handed antiquark, so e.g. the amplitude 
$g_R(-p_a) q_R(-p_b)\to g_R(p_1) g_R(p_2) q_R(p_{b'})$ becomes $\cM(a^-_g, 1^+_g, 2^+_g, {b'}^+_q, b^-_{\bar q})$.}
\begin{equation}
C^{q(0)}(p_b^-;p_{b'}^+) =
i \sqrt{p_{b'\perp}^* \over p_{b'\perp}}\, ,\label{centrcqq}
\end{equation} 
which under parity becomes~\cite{DelDuca:1999iql}
\begin{equation}
[C^{q(0)}(p_b^{-\nu};p_{b'}^\nu) ]^\ast = S C^{q(0)}(p_b^\nu;p_{b'}^{-\nu}) \qquad {\rm with} \quad
S = - {\rm sign} ({\bar q}^0 q^0 ) \,.
\end{equation} 
If Regge factorisation holds at NLL accuracy, then we can write the two-quark three-gluon amplitude in the NMRK region of \eqn{qmrapp1} as
%
%
\beq
\left. \cM^{(-)[8_a]}_{gq\to ggq}\right|_{NMRK}
= \sum_{\sigma \in S_2}\gs^3\, (F^{d_{\sigma_1}}F^{d_{\sigma_2}})_{ac} (T^c)_{b'\bar{b}} \,
 \left. m_{2q3g}(p_a^{\nu_a}, p^{\nu_{\sigma_1}}_{\sigma_1}, p^{\nu_{\sigma_2}}_{\sigma_2}, p_{b'}^{\nu_{b'}}, p_b^{\nu_b})\right|_{NMRK} \,,
\label{NLLfactorizationqq}
\eeq
with the colour-ordered amplitude,
\beqa
\lefteqn{ \left. m_{2q3g}(p_a^{\nu_a}, p^{\nu_1}_1, p^{\nu_2}_2, p_{b'}^{\nu_{b'}}, p_b^{\nu_b})\right|_{NMRK} } \nn\\
&=& \frac1{2} {s\over t} A^{gg}(p_a^{\nu_a}; p^{\nu_1}_1, p^{\nu_2}_2) 
\left[ \left({s_{2 b'}\over \tau}\right)^{\alpha(t)} + \left({-s_{2 b'}\over \tau}\right)^{\alpha(t)} \right] 
C^{q}(p_b^{\nu_b};p_{b'}^{\nu_{b'}}) \,,  
\label{NLLfactorizationqq2}
\eeqa
where $A^{gg}(p_a^{\nu_a}; p^{\nu_1}_1, p^{\nu_2}_2 )$ is given by \eqn{eq:a1loop}, 
and $C^{q}(p_b^{\nu_b};p_{b'}^{\nu_{b'}})$ is given by the quark analog of \eqn{fullv}.
We can then write the colour-ordered amplitude (\ref{NLLfactorizationqq2}) 
as a double expansion in the strong coupling $\as$ and in $\log(s_{2 b'}/\tau)$, 
and repeat the steps of \eqn{nllnmrkexpand} through \eqn{eq:1loopsubtr}.
Once expanded at one loop, the real part of \eqn{NLLfactorizationqq2} becomes
\beq
\RE\left[ m^{(1)}_{2q3g} \right]_{NMRK} 
= \alpha^{(1)}(t) \log\left(\frac{s_{\sigma_2 b'}}{\tau}\right) + \RE\big[ A^{gg(1)}(p_{\sigma_1}, p_{\sigma_2},\tau) \big] 
+ \RE\big[ C^{q (1)}(t,\tau) \big] \,,
\label{eq:1loopsubtrq}
\eeq
with $A^{gg(1)}$ as in \eqn{oneloopIFgg}, and where,
through $\ord(\eps^0)$, the $\overline{\rm MS}$-renormalised one-loop quark impact factor 
$C^{q(1)}(t)$ can be written as~\cite{DelDuca:1998kx,DelDuca:2017pmn}
\beqa
\RE\big[ C^{q(1)}_{-+}(t,\tau) \big] &=& \cg \left\{ \left({\mu^2\over -t}\right)^{\epsilon} 
\left[ - \frac{\gamma_K^{(1)} }{\epsilon^2} C_F + \frac4{\eps} \gamma_q^{(1)} + \frac{\beta_0}{2\eps} + \frac{N_c}{\eps} \log\left(\frac{\tau}{-t}\right)
\right. \right. \nn\\
&& \left.\left. \qquad\qquad + \left( {19\over 18} - {\delta_R\over 3} + {\pi^2\over 2} \right) N_c - \frac5{9} N_f 
+ \left( {7\over 2} + {\delta_R\over 2}\right) {1\over N_c}  \right] - \frac{\beta_0}{2\eps} \right\} \nn\\
&=& \cg \left\{ \left({\mu^2\over -t}\right)^{\epsilon}  
\left[ \left(-{1\over\epsilon^2} + {1\over 3\epsilon} + \frac1{\eps} \log\left(\frac{\tau}{-t}\right) + {19\over 18} - {\delta_R\over 3} + {\pi^2\over 2} \right) N_c
\right.\right. \nonumber\\ & &\qquad \left.\left.
- \left( {1\over 3\epsilon} + {5\over 9}\right) N_f
 + \left({1\over\epsilon^2} + {3\over 2\epsilon} + {7\over 2} + {\delta_R\over 2}\right) {1\over N_c} 
 \right] - {\beta_0\over 2\epsilon} \right\} \,,
\label{1loopqifeps}
\eeqa
with $\gamma_q^{(1)}$ the one-loop coefficient of the quark collinear anomalous dimension (\ref{eq:c1}).

In order to verify that the one-loop two-quark three-gluon amplitude in NMRK takes the factorised form given by \eqn{eq:1loopsubtrq},
we use the colour decomposition of the one-loop two-quark three-gluon amplitude~\cite{DelDuca:1999rs},
\beqa
\lefteqn{ M_{2q3g}^{1-loop}(1_{\bar{q}},2_q ,3,4,5) } \nn\\
&=& \gs^5 \left[ \sum_{p=2}^5 \sum_{\sigma \in S_3} \left( T^{c_2}T^{d_{\sigma_3}}\ldots T^{d_{\sigma_p}} T^{c_1} \right)_{i_2}^{\,\,\bar{i}_1} 
\left( F^{d_{\sigma_{p+1}}} \ldots F^{d_{\sigma_5}}\right)_{c_1c_2}  \right. \nn\\
&& \qquad\times A_5^{R, [1]}(1_{\bar{q}},\sigma_{p+1},\ldots,\sigma_5, 2_q, \sigma _3,\ldots,\sigma_p ) \nn\\
&&\quad + \left. \frac{N_f}{N_c} \sum_{j=1}^{4} \sum_{\sigma \in S_3 / S_{5;j}} Gr_{5;j}^{\bar{q} q} (\sigma_3,\sigma_4, \sigma_5 ) A_{5;j}^{[1/2]} (1_{\bar{q}}, 2_q, \sigma_3, \sigma_4, \sigma_5 )    \right]\,,
\label{basiceq}
\eeqa
where the colour structures $Gr_{5;j}^{\bar{q} q}$ are defined by 
\beqa
Gr_{5;1}^{\bar{q} q}(3, 4, 5) &=& N_c \left( T^{d_3} T^{d_4} T^{d_5} \right)_{i_2}^{\,\,\bar{i}_1}\,,  \nn\\
Gr_{5;2}^{\bar{q} q}(3; 4, 5) &=& 0 \,, \nn\\
Gr_{5;3}^{\bar{q} q}(3, 4; 5) &=& \Tr \left( T^{d_3} T^{d_4} \right) \left( T^{d_5} \right)_{i_2}^{\,\,\bar{i}_1} \,,\nn\\
Gr_{5;4}^{\bar{q} q}(3, 4; 5) &=& \Tr \left( T^{d_3} T^{d_4} T^{d_5} \right) \delta_{i_2}^{\,\,\bar{i}_1} \,,
\label{eq:gr5}
\eeqa
and $S_{5;j} = \mathbb{Z}_{j-1}$ is the subgroup of $S_3$ that leaves $Gr_{5;j}^{\bar{q} q}$ invariant, and
\beqa
A_{5;1}^{[1/2]} (1_{\bar{q}}, 2_q, 3, 4, 5 ) &=& A_{5}^{L,[1/2]} (1_{\bar{q}}, 2_q, 3, 4, 5 ) \,,\nn\\
A_{5;3}^{[1/2]} (1_{\bar{q}}, 2_q; 3, 4; 5 ) &=& - \sum_{\sigma \in \{4,3\} \dsqcup \{1,2,5\} } A_{5;1}^{[1/2]} \left( \sigma(1_{\bar{q}}, 2_q, 3, 4, 5 ) \right)\,,\nn\\
A_{5;4}^{[1/2]} (1_{\bar{q}}, 2_q; 3, 4; 5 ) &=& \sum_{\sigma \in \{5,4,3\} \dsqcup \{1,2\} } A_{5;1}^{[1/2]} \left( \sigma(1_{\bar{q}}, 2_q, 3, 4, 5 ) \right)\,,
\label{eq:a51/2}
\eeqa
where the shuffle product $\{\alpha\}\dsqcup \{\beta\}$ yields the permutations which preserve the orderings of the elements in $\{\alpha\}$
and in $ \{\beta\}$ while allowing for all possible orderings of the elements of $\{\alpha\}$ with respect to $ \{\beta\}$. In \eqn{eq:a51/2}, we did not list
the $j=2$ case, since the corresponding colour structure is missing in \eqn{eq:gr5}.
Further, we spell out the first sum of \eqn{basiceq},
\beqa
\lefteqn{ \sum_{p=2}^5 \sum_{\sigma \in S_3} \left( T^{c_2}T^{d_{\sigma_3}}\ldots T^{d_{\sigma_p}} T^{c_1} \right)_{i_2}^{\,\,\bar{i}_1} 
\left( F^{d_{\sigma_{p+1}}} \ldots F^{d_{\sigma_5}}\right)_{c_1c_2} } \nn\\
&& \qquad\times A_5^{R, [1]}(1_{\bar{q}},\sigma_{p+1},\ldots,\sigma_5, 2_q, \sigma _3,\ldots,\sigma_p ) \nn\\
&=&  \sum_{\sigma \in S_3} \left[ \left( T^{c_2} T^{c_1} \right)_{i_2}^{\,\,\bar{i}_1} 
\left( F^{d_{\sigma_{3}}} F^{d_{\sigma_{4}}} F^{d_{\sigma_5}}\right)_{c_1c_2} 
 A_5^{R, [1]}(1_{\bar{q}},\sigma_{3},\sigma _4,\sigma_5, 2_q ) \right. \nn\\
&& \qquad + \left( T^{c_2} T^{d_{\sigma_{3}}} T^{c_1} \right)_{i_2}^{\,\,\bar{i}_1} 
\left( F^{d_{\sigma_{4}}} F^{d_{\sigma_5}}\right)_{c_1c_2}  A_5^{R, [1]}(1_{\bar{q}},\sigma _4,\sigma_5, 2_q,\sigma_{3} ) \nn\\
&& \qquad + \left( T^{c_2} T^{d_{\sigma_{3}}} T^{d_{\sigma_{4}}} T^{c_1} \right)_{i_2}^{\,\,\bar{i}_1} 
\left( F^{d_{\sigma_5}}\right)_{c_1c_2}  A_5^{R, [1]}(1_{\bar{q}},\sigma_5, 2_q,\sigma_{3},\sigma _4 ) \nn\\
&& \left. \qquad + \left( T^{c_2} T^{d_{\sigma_{3}}} T^{d_{\sigma_{4}}} T^{d_{\sigma_5}} T^{c_1} \right)_{i_2}^{\,\,\bar{i}_1} 
\delta_{c_1c_2}  A_5^{R, [1]}(1_{\bar{q}}, 2_q,\sigma_{3},\sigma _4,\sigma_5 ) \right] \,.
\label{basiceq2}
\eeqa
Finally, the reflection identity,
\beq
A_n^{R, [1]}(1_{\bar{q}},3,\ldots,2_q, \ldots, n-1,n ) = (-1)^n A_n^{L, [1]}(1_{\bar{q}}, n, n-1,\ldots, 2_q, \ldots, 3)\,.
\label{eq:reflid}
\eeq
relates the left and right primitive amplitudes.

The primitive amplitudes $A_5^{L, [1]}(1_{\bar{q}}, 2_q,3,4,5 )$ and $A_{5}^{L,[1/2]} (1_{\bar{q}}, 2_q, 3, 4, 5 )$, which are necessary to assemble
the one-loop two-quark three-gluon amplitude (\ref{basiceq}), have been provided by Ref.~\cite{Bern:1994fz} in the DR/FDH schemes. 
We have evaluated them
in the NMRK region of \eqn{qmrapp1} and \app{sec:appc}. In NMRK, the leading contributions are given by the terms with $p=2$, $p=5$ and $j=1$ in
\eqn{basiceq}, i.e. by the first and the last term on the right-hand side of \eqn{basiceq2} and by the first term of \eqns{eq:gr5}{eq:a51/2}.

\subsubsection{The $p=2$ term}

The knowledge of the primitive amplitudes $A_5^{L, [1]}(1_{\bar{q}}, 2_q,3,4,5 )$~\cite{Bern:1994fz}, and the reflection identity (\ref{eq:reflid})
allow us to compute the primitive amplitudes $A_5^{R, [1]}$, which, in the notation of the $g_a\, q_b\to g_1\, g_2\,q_{b'}$ scattering, are
$A_5^{R, [1]}(b_{\bar{q}},\sigma_{a},\sigma _1,\sigma_2, {b'}_q )$. Computing the primitive amplitudes $A_5^{R, [1]}$
in the NMRK of \eqn{qmrapp1} for different colour orderings, we find that they are related by,
\beqa
A_5^{R, [1]}(b_{\bar{q}},a,1, 2, {b'}_q ) &=& A_5^{R, [1]}(b_{\bar{q}}, 2, 1, a, {b'}_q ) = 
- \lambda_1 m_{2q3g}^{(0)}(p_a, p_1, p_2, p_{b'}, p_b ) \,,\nn\\
A_5^{R, [1]}(b_{\bar{q}}, a, 2, 1, {b'}_q ) &=& A_5^{R, [1]}(b_{\bar{q}}, 1, 2, a, {b'}_q ) =
- \lambda_2 m_{2q3g}^{(0)}(p_a, p_2, p_1, p_{b'}, p_b ) \,,\nn\\
A_5^{R, [1]}(b_{\bar{q}}, 1, a, 2, {b'}_q ) &=& A_5^{R, [1]}(b_{\bar{q}}, 2, a,1, {b'}_q ) \nn\\
&=& \lambda_1 m_{2q3g}^{(0)}(p_a, p_1, p_2, p_{b'}, p_b ) + \lambda_2 m_{2q3g}^{(0)}(p_a, p_2, p_1, p_{b'}, p_b )\,,
\label{eq:a12qq}
\eeqa
with $\lambda_1, \lambda_2$ in \eqns{eq:g1}{eq:g2}, and $m_{2q3g}^{(0)}(p_a, p_2, p_1, p_{b'}, p_b )$ the tree colour-ordered amplitude
(\ref{NLOfactorizationqq2}), where a $NMRK$ subscript is understood.
Using \eqn{eq:a12qq}, the colour coefficient of the $p=2$ term of \eqn{basiceq},
i.e. of the first term on the right-hand side of \eqn{basiceq2}, simplifies and factors an $N_c$ term out. It can be written as
\beqa
&& \sum_{\sigma \in S_3} \left( T^{c_2} T^{c_1} \right)_{b'}^{\,\,\bar{b}} 
\left( F^{d_{\sigma_{a}}} F^{d_{\sigma_{1}}} F^{d_{\sigma_2}}\right)_{c_1c_2} 
 A_5^{R, [1]}(b_{\bar{q}},\sigma_{a},\sigma _1,\sigma_2, {b'}_q ) \nn\\
&&  \stackrel{\scriptscriptstyle NMRK}{\Longrightarrow}
N_c  \sum_{\sigma \in S_2} (F^{d_{\sigma_1}}F^{d_{\sigma_2}} )_{ac} \, (T^c)_{b'}^{\bar{b}} \lambda_{\sigma_1}
m_{2q3g}^{(0)}(p_a, p_{\sigma_1}, p_{\sigma_2}, p_{b'}, p_b ) \,.
\label{eq:ncnmrk}
\eeqa
Thus, the $p=2$ term provides the leading $N_c$ terms of the one-loop two-quark three-gluon amplitude in NMRK.

\subsubsection{The $p=5$ term}

For the $p=5$ term, one can use the relation~\cite{Bern:1994fz},
\beqa
\lefteqn{ A_5^{R, [1]}(b_{\bar{q}},{b'}_q, \sigma_{a},\sigma _1,\sigma_2) } \\
&=& A_5^{SUSY}(b_{\bar{q}},{b'}_q, \sigma_{a},\sigma _1,\sigma_2) 
- A_5^{L, [1]}(b_{\bar{q}},{b'}_q, \sigma_{a},\sigma _1,\sigma_2)  - A_5^{L, [1/2]}(b_{\bar{q}},{b'}_q, \sigma_{a},\sigma _1,\sigma_2)\,, \nn
\eeqa
where the primitive amplitudes on the right-hand side are all provided in Ref.~\cite{Bern:1994fz}. Computing the primitive amplitudes $A_5^{R, [1]}$
in the NMRK of \eqn{qmrapp1} for different colour orderings, we find that they are related by
\beqa
A_5^{R, [1]}(b_{\bar{q}} , b'_q, a, 1, 2) &=& A_5^{R, [1]}(b_{\bar{q}} , b'_q, 2, 1, a) = \Phi\, m_{2q3g}^{(0)}(p_a, p_1, p_2, p_b, p_{b'} )\,, \nn\\
A_5^{R, [1]}(b_{\bar{q}} , b'_q, a, 2, 1) &=& A_5^{R, [1]}(b_{\bar{q}} , b'_q, 1, 2, a) = \Phi\, m_{2q3g}^{(0)}(p_a, p_2, p_1, p_b, p_{b'} )\,,  \nn\\
A_5^{R, [1]}(b_{\bar{q}} , b'_q, 1, a, 2) &=& A_5^{R, [1]}(b_{\bar{q}} , b'_q, 2, a, 1) \nn\\
&=& - \Phi \left[ m_{2q3g}^{(0)}(p_a, p_1, p_2, p_b, p_{b'} ) + m_{2q3g}^{(0)}(p_a, p_2, p_1, p_b, p_{b'} )\right],
\label{eq:qqa12}
\eeqa
where $m_{2q3g}^{(0)}(p_a, p_1, p_2, p_b, p_{b'} ) = - m_{2q3g}^{(0)}(p_a, p_1, p_2, p_{b'}, p_b )$, and with
\beq
\Phi = \cg \left( \frac{\mu^2}{-t} \right)^\epsilon \left( \frac{1}{\epsilon^2} + \frac{3}{2 \epsilon} + \frac{7}{2} \right)  \,.
\eeq
Using \eqn{eq:qqa12}, the colour coefficient of the $p=5$ term of \eqn{basiceq},
i.e. of the fourth term on the right-hand side of \eqn{basiceq2}, simplifies and factors a $1/N_c$ term out. It can be written as
\beqa
&& \sum_{\sigma \in S_3} \delta_{c_1c_2}
\left( T^{c_2} T^{d_{\sigma_{a}}} T^{d_{\sigma_{1}}} T^{d_{\sigma_2}} T^{c_1} \right)_{b'}^{\,\,\bar{b}} 
  A_5^{R, [1]}(b_{\bar{q}}, b'_q,\sigma_a,\sigma _1,\sigma_2 ) \nn\\
&&  \stackrel{\scriptscriptstyle NMRK}{\Longrightarrow}
\frac1{N_c} \Phi \sum_{\sigma \in S_2} (F^{d_{\sigma_1}}F^{d_{\sigma_2}} )_{ac} \, (T^c)_{b'}^{\bar{b}} \lambda_{\sigma_1}
m_{2q3g}^{(0)}(p_a, p_{\sigma_1}, p_{\sigma_2}, p_b, p_{b'} ) \,.
\label{eq:1ncnmrk}
\eeqa
Thus, the $p=5$ term provides the $1/N_c$ terms of the one-loop two-quark three-gluon amplitude in NMRK.

\subsubsection{The $j=1$ term}

For the $j=1$ term, we use the first term of \eqns{eq:gr5}{eq:a51/2}. Computing the primitive amplitudes $A^{[1/2]}_{5;1}$
in the NMRK of \eqn{qmrapp1} for different colour orderings, we find that they are related by,
\beqa
A^{[1/2]}_{5;1}(b_{\bar{q}} , b'_q, a, 1, 2) &=& A^{[1/2]}_{5;1}(b_{\bar{q}} , b'_q, 2, 1, a) = \chi_1\, m_{2q3g}^{(0)}(p_a, p_1, p_2, p_b, p_{b'} )\,, \nn\\
A^{[1/2]}_{5;1}(b_{\bar{q}} , b'_q, a, 2, 1) &=& A^{[1/2]}_{5;1}(b_{\bar{q}} , b'_q, 1, 2, a) = \chi_2\, m_{2q3g}^{(0)}(p_a, p_2, p_1, p_b, p_{b'} )\,, \nn\\
A^{[1/2]}_{5;1}(b_{\bar{q}} , b'_q, 1, a, 2) &=& A^{[1/2]}_{5;1}(b_{\bar{q}} , b'_q, 2, a, 1) \nn\\
&=& -\chi_1 m_{2q3g}^{(0)}(p_a, p_1, p_2, p_b, p_{b'} ) - \chi_2 m_{2q3g}^{(0)}(p_a, p_2, p_1, p_b, p_{b'} )\,,
\label{eq:1/2bba12}
\eeqa
with $\chi_1, \chi_2$ in \eqns{eq:g3}{eq:g4}.
Using \eqn{eq:1/2bba12}, the colour coefficient of the $j=1$ term of \eqn{basiceq} simplifies and can be written as
\beqa
&& N_f \sum_{\sigma \in S_3}  \left( T^{d_{\sigma_{a}}} T^{d_{\sigma_{1}}} T^{d_{\sigma_2}} \right)_{b'}^{\,\,\bar{b}}  
A^{[1/2]}_{5;1}(b_{\bar{q}} , b'_q, \sigma_a,\sigma _1,\sigma_2) \nn\\
&&  \stackrel{\scriptscriptstyle NMRK}{\Longrightarrow} 
N_f \sum_{\sigma \in S_2} (F^{d_{\sigma_1}}F^{d_{\sigma_2}} )_{ac} \, (T^c)_{b'}^{\bar{b}} \chi_{\sigma_1}
m_{2q3g}^{(0)}(p_a, p_{\sigma_1}, p_{\sigma_2}, p_b, p_{b'} )\,.
\label{eq:nfnmrk}
\eeqa
Thus, the $j=1$ term provides the $N_f$ terms of the one-loop two-quark three-gluon amplitude in NMRK.

\subsubsection{Regge factorisation}

Combining \eqns{eq:ncnmrk}{eq:1ncnmrk} and (\ref{eq:nfnmrk}), we obtain the one-loop two-quark three-gluon 
amplitude in the NMRK region of \eqn{qmrapp1},
\beqa
\lefteqn{ \RE\left[ \cM^{(1)}_{gq\to ggq} \right]_{NMRK} } \nn\\
&=& \sum_{\sigma \in S_2} (F^{d_{\sigma_1}}F^{d_{\sigma_2}} )_{ac} \, (T^c)_{b'}^{\bar{b}} 
\left( N_c \lambda_{\sigma_1} + N_f \chi_{\sigma_1} + \frac{\Phi}{N_c} \right) m_{2q3g}^{(0)}(p_a, p_{\sigma_1}, p_{\sigma_2}, p_{b'}, p_b ) \,.
\eeqa
Then we use  \eqn{eq:1loopsubtrq}, with the one-loop Regge trajectory (\ref{alph}) and
the one-loop quark impact factor (\ref{1loopqifeps}), which in particular subtracts out all the $1/N_c$ terms,
and we obtain the one-loop impact factor for the emission of two gluons $A^{gg(1)}$,
which is in agreement with \eqn{oneloopIFgg}, thus verifying that Regge factorisation holds at NLL accuracy in NMRK.

\subsection{Helicity-violating contributions}
\label{sec:helviol1loopif}

The conservation of helicity along the $s$-channel direction in Minkowski space, that we mentioned in the Introduction, 
is an approximate feature of the Regge limit of tree four-gluon amplitudes, since the helicity-flip impact factor $C^{g(0)}(p^+;p'^+)$ exists, 
but is power suppressed in $t/s$. At one loop, the four-gluon amplitude $m_4^{(1)}(-,+,+,+)$ is not vanishing, so the
helicity-flip impact factor $C^{g(1)}(p^+;p'^+)$ is not power suppressed in $t/s$~\cite{Fadin:1993wh,Fadin:1992zt,DelDuca:1998kx}.
In fact, we can write
\begin{equation}
C^{g}(p_j^+;p_{j'}^+) = 
C^{g (0)}(p_j^-;p_{j'}^+)\left(\frac{\as}{4\pi} C^{g (1)}_{++}(q_\perp) + \ord(\as^2) \right)\,, 
\label{eq:helviol}
\end{equation}
with $j=a, b$ and $q_\perp = p_{b'\perp} = - p_{a'\perp}$, and
\beq
C^{g (1)}_{++}(q_\perp) = - \frac{N_c-N_f}3\, \frac{q_\perp^\ast}{q_\perp}\,.
\label{eq:1lhelviol}
\eeq
Because the tree four-gluon amplitude $m_4^{(0)}(-,+,+,+)$ vanishes, the one loop four-gluon amplitude $m_4^{(1)}(-,+,+,+)$
contributes to a squared amplitude, and so to a cross section, only when it is squared by itself.
Thus, although the helicity-flip impact factor $C^{g}(p^+;p'^+)$ contributes to the four-gluon amplitude at NLL accuracy,
it occurs in a squared amplitude only at NNLL accuracy. As such, it will contribute to the computation of the jet impact factor at NNLO in $\as$, 
mentioned at the end of \sec{sec:reggennll}.

Likewise, although the tree five-gluon amplitude $m_5^{(0)}(-,+,+,+,+)$ vanishes, the one loop amplitude $m_5^{(1)}(-,+,+,+,+)$ does not.
In general kinematics, its expression is given in Ref.~\cite{Bern:1993mq}. In the NMRK region of \eqn{qmrapp1}, it becomes
\beqa
\lefteqn{ M_{5:1}^{[1]}(b^-,a^+,1^+,2^+,{b'}^+) } \nn\\ &=& \frac{\gs^2}{48\pi^2} \left(N_c-N_f\right) M_5^{(0)}(b^-,a^-,1^+,2^+,{b'}^+) 
\left( - x_2 \frac{p_{b'\perp}^\ast}{p_{b'\perp}} - x_1 \frac{p_{1\perp}^\ast}{p_{1\perp}}
+ x_1x_2 \frac{[12]^2}{s_{12}} \right) \,,
\label{eq:m51helviol}
\eeqa
with $x_{1,2}$ in \eqn{sec:x12}. From \eqn{eq:m51helviol} we may derive the helicity-flip impact factor $A^{gg}(p_a^+; p^+_1, p^+_2 )$
for the emission of two gluons,
\beq
A^{gg}(p_a^+; p^+_1, p^+_2) = 
A^{gg(0)}(p_a^-; p^+_1, p^+_2 ) \left(\frac{\as}{4\pi} \bar{A}^{gg(1)}_+(p_1,p_2) + \ord(\as^2) \right)\, ,
\label{eq:a1loophelviol}
\eeq
with
\beq
\bar{A}^{gg(1)}_+(p_1,p_2) = \frac{N_c-N_f}3\, \left( - x_2 \frac{q_\perp^\ast}{q_\perp} - x_1 \frac{p_{1\perp}^\ast}{p_{1\perp}}
+ x_1x_2 \frac{[12]^2}{s_{12}} \right) \,,
\label{eq:a1loopifhelviol}
\eeq
with $q_\perp = p_{b'\perp} = - (p_{1\perp} + p_{2\perp})$. 
Note however that although the helicity-flip impact factor $A^{gg}(p_a^+; p^+_1, p^+_2 )$ contributes to the five-gluon amplitude at NLL accuracy,
it contributes to the jet impact factor only at next-to-next-to-next-to-leading order ($\mathrm{N^3LO}$) in $\as$.

\section{Discussion and conclusions}
\label{sec:conc}

In the Regge limit of QCD amplitudes, large logarithms $\log(s/t)$ arise.
For octet exchange in the $t$ channel, they are resummed by the BFKL equation at leading logarithmic and NLL accuracy.
Also the other colour representations, which are exchanged in the $t$ channel of gluon-gluon scattering amplitudes, have been fully
analysed at NLL accuracy~\cite{Caron-Huot:2017fxr,Caron-Huot:2013fea,Caron-Huot:2017zfo,Caron-Huot:2020grv}.
Moving toward a BFKL equation at NNLL accuracy presents several challenges, which go from taking stock of the exchange 
of three Reggeised gluons~\cite{Caron-Huot:2017fxr,Fadin:2016wso,Fadin:2017nka,Falcioni:2020lvv}, which violate Regge-pole
factorisation~\cite{DelDuca:2001gu}, to computing loop-level building blocks of the BFKL kernel in multi-Regge and
next-to-multi-Regge kinematics.

In this paper, we have evaluated the one-loop impact factors for the emission of two gluons, \eqns{oneloopIFgg}{eq:a1loopifhelviol},
by taking the one-loop five-gluon~\cite{Bern:1993mq} and two-quark three-gluon amplitudes~\cite{Bern:1994fz} in NMRK.
As a sanity check, we verified that in MRK, \eqn{oneloop1} is reduced to the one-loop five-gluon amplitude which yields the one-loop central-emission vertex~\cite{DelDuca:1998cx}. 

The form of \eqn{oneloopIFgg} depends on the reggeisation ansatz (\ref{NLLfactorization}).
In \eqn{NLLfactorization}, we have chosen a different base of the Regge-trajectory exponent for each of the two
colour-ordered amplitudes. Such issue is immaterial in the Regge limit (\ref{sudall}) or in MRK~\cite{DelDuca:1998cx}, 
where there is only one colour-ordered amplitude.
The ansatz (\ref{NLLfactorization}) ensures that in the MRK limit the impact factor (\ref{oneloopIFgg}) has the correct factorisation properties,
i.e. that it is reduced to \eqn{eq:mrklipvert}.
We note that in \eqn{NLLfactorization} one could avoid introducing a different base of the trajectory exponent for each
colour-ordered amplitude, by taking as a base the geometric $\frac{\sqrt{s_{1b'}s_{2b'}}}{\tau}$ or the arithmetic $\frac{s_{1b'}+s_{2b'}}{2\tau}$
means of $s_{1b'}$ and $s_{2b'}$. That would induce terms of the type $\log\left(\frac{s_{2b'}}{s_{1b'}}\right)$ or 
$\log\left(\frac{s_{2b'}}{s_{1b'}+s_{2b'}}\right)$ in the impact factor (\ref{oneloopIFgg}). Such terms would be harmless in NMRK, but would not
yield the correct MRK limit (\ref{eq:mrklipvert}).

\Eqn{oneloopIFgg} is the last ingredient which is necessary to evaluate the gluon-jet impact factor at NNLO accuracy in $\as$.
It is also the first instance in which loop-level QCD amplitudes are evaluated in 
next-to-multi-Regge kinematics\footnote{Loop-level ${\cal N}=4$ SYM amplitudes have been considered in NMRK in order to compute 
the two-loop six-point remainder function~\cite{DelDuca:2009au,DelDuca:2010zg}.}. 
This has led to the introduction of a specific reggeisation ansatz for each colour-ordered amplitude (\ref{NLLfactorization}).
We expect that the same issue will arise in
the evaluation of the one-loop corrections to the central-emission vertex for the production of two gluons or 
of a quark-antiquark pair along the gluon ladder, which is one of the building blocks, yet to be determined, of the 
BFKL kernel at NNLL accuracy.

\section*{Acknowledgements}

We thank Einan Gardi for useful discussions and for a critical reading of the draft. VDD's work has been supported in part by the
European Research Council (ERC) under grant agreement No 694712 (PertQCD).
MC's work has been supported in part by the European Research Council under Advanced Investigator Grant ERC-AdG-885414.

\appendix

\section{Multiparton kinematics}
\label{sec:appa}

Without loss of generality,
we consider the production of up to three partons of momenta $p_1$, $p_2$
and $p_3$, in the scattering between two partons of  
momenta $p_a$ and $p_b$. By convention,
we consider the scattering in the unphysical region where all momenta 
are taken as outgoing, and then we analitically continue to the
physical region where $p_a^0<0$ and $p_b^0<0$. Thus
partons are incoming or outgoing depending on the sign
of their energy. Since the helicity of a positive-energy 
(negative-energy) massless spinor has the same (opposite) sign as its
chirality, the helicities assigned to the partons 
depend on whether they are incoming or outgoing.
We label outgoing (positive-energy) particles 
with their helicity; if they are incoming the 
actual helicity and charge are reversed, e.g. an incoming left-handed
parton is labelled as an outgoing right-handed anti-parton.

Using light-cone coordinates $p^{\pm}= p_0\pm p_z $, and
complex transverse coordinates $p_{\perp} = p^x + i p^y$, with scalar
product,
\begin{equation}
2 p\cdot q = p^+q^- + p^-q^+ - p_{\perp} q^*_{\perp} - p^*_{\perp} q_{\perp}\,, 
\label{eq:scalprod}
\end{equation} 
the four-momenta are,
\begin{eqnarray}
p_a &=& \left(p_a^+/2, 0, 0,p_a^+/2 \right) 
     \equiv  \left(p_a^+ , 0; 0, 0 \right)\, ,\nonumber \\
p_b &=& \left(p_b^-/2, 0, 0,-p_b^-/2 \right) 
     \equiv  \left(0, p_b^-; 0, 0\right)\, ,\label{in}\\
p_i &=& \left( (p_i^+ + p_i^- )/2, 
                {\rm Re}[p_{i\perp}],
                {\rm Im}[p_{i\perp}], 
                (p_i^+ - p_i^- )/2 \right)\nonumber\\
    &\equiv& \left(|p_{i\perp}| e^{y_i}, |p_{i\perp}| e^{-y_i}; 
|p_{i\perp}|\cos{\phi_i}, |p_{i\perp}|\sin{\phi_i}\right)\, \,,\nonumber
\end{eqnarray}
where $y$ is the rapidity, and $1\le i \le 3$. The first notation in \eqn{in} is the 
standard representation 
$p^\mu =(p^0,p^x,p^y,p^z)$, while in the second we have the $+$ and $-$
light-cone components on the left of the semicolon, 
and on the right the transverse components.

From the momentum conservation,
\begin{eqnarray}
0 &=& \sum_{i=1}^3 p_{i\perp}\, ,\nonumber \\
p_a^+ &=& -\sum_{i=1}^3 p_i^+\, ,\label{nkin}\\ 
p_b^- &=& -\sum_{i=1}^3 p_i^-\, ,\nonumber
\end{eqnarray}
and using the scalar product \eqn{eq:scalprod},
the Mandelstam invariants may be written as,
\begin{eqnarray}
s &=& 2 p_a\cdot p_b = \sum_{i,j=1}^3 p_i^+ p_j^- \nonumber\\ 
s_{ai} &=& 2 p_a\cdot p_i = -\sum_{j=1}^3 p_i^- p_j^+ \label{inv}\\ 
s_{bi} &=& 2 p_b\cdot p_i = -\sum_{j=1}^3 p_i^+ p_j^- \nonumber.
\end{eqnarray}
For the momenta (\ref{in}), we use the spinor products~\cite{DelDuca:1999iql}
\begin{eqnarray}
\langle p_i p_j\rangle &=& p_{i\perp}\sqrt{p_j^+\over p_i^+} - p_{j\perp}
\sqrt{p_i^+\over p_j^+}\, , \nonumber\\ 
\langle p_a p_i\rangle &=& - i \sqrt{-p_a^+
\over p_i^+}\, p_{i\perp}\, ,\label{spro}\\ 
\langle p_i p_b\rangle &=&
i \sqrt{-p_b^- p_i^+}\, ,\nonumber\\ 
\langle p_a p_b\rangle 
&=& -\sqrt{p_a^+p_b^-}\, ,\nonumber
\end{eqnarray}
where we have used the mass-shell condition 
$|p_{i\perp}|^2 = p_i^+ p_i^-$.

\section{Multi-Regge kinematics}
\label{sec:appb}

In the multi-Regge kinematics, we require that the gluons
are strongly ordered in rapidity and have comparable transverse momentum,
\begin{equation}
y_1 \gg y_2 \gg y_3;\qquad |p_{1\perp}| \simeq |p_{2\perp}| \simeq|p_{3\perp}|\, 
.\nonumber
\end{equation}
Momentum conservation (\ref{nkin}) then becomes
\begin{eqnarray}
0 &=& \sum_{i=1}^3 p_{i\perp}\, ,\nonumber \\
p_a^+ &\simeq& -p_1^+\, ,\label{mrkin}\\ 
p_b^- &\simeq& -p_3^-\, .\nonumber
\end{eqnarray}
The Mandelstam invariants (\ref{inv}) are reduced to,
\begin{eqnarray}
s &=& 2 p_a\cdot p_b \simeq p_1^+ p_3^-\,, \nonumber\\ 
s_{ai} &=& 2 p_a\cdot p_i \simeq - p_1^+ p_i^- \,,\label{mrinv}\\ 
s_{bi} &=& 2 p_b\cdot p_i \simeq - p_i^+ p_3^-\,, \nonumber\\ 
s_{ij} &=& 2 p_i\cdot p_j \simeq p_i^+ p_j^-  \qquad {\rm for}\, y_i>y_j\,,
\nonumber
\end{eqnarray}
to leading accuracy. The spinor products (\ref{spro}) become,
\begin{eqnarray}
\langle p_i p_j\rangle &\simeq& -\sqrt{p_i^+\over p_j^+}\,
p_{j\perp}\, \qquad {\rm for}\, y_i>y_j \;, \nonumber\\
\langle p_a p_i\rangle &\simeq& - i\sqrt{p_1^+\over p_i^+}\,
p_{i\perp}\, ,\label{mrpro}\\ \langle p_i p_b\rangle 
&\simeq& i\sqrt{p_i^+ p_3^-}\, ,\nonumber\\ 
\langle p_a p_b\rangle &\simeq& -\sqrt{p_1^+ p_3^-}\, .\nonumber
\end{eqnarray}

\section{Next-to-multi-Regge kinematics}
\label{sec:appc}

We consider the production of three partons of momenta $p_1,p_2,p_3$,
with partons 1 and 2 in the forward-rapidity region of parton $p_a$,
\begin{equation}
y_1 \simeq y_2 \gg y_3\,;\qquad |p_{1\perp}|
\simeq |p_{2\perp}| \simeq |p_{3\perp}|\, .\label{qmrapp}
\end{equation}
Momentum conservation (\ref{nkin}) becomes
\begin{eqnarray}
0 &=& \sum_{i=1}^3 p_{i\perp}\, ,\nonumber \\
p_a^+ &\simeq& -(p_1^+ + p_2^+)\, ,\label{frkapp}\\ 
p_b^- &\simeq& -p_3^-\, .\nonumber
\end{eqnarray}
The Mandelstam invariants (\ref{inv}) become
\begin{eqnarray}
s &\simeq& (p_1^+ + p_2^+) p_3^- \nonumber\\ 
s_{a3} &\simeq& - (p_1^+ + p_2^+) p_3^- \nn\\
s_{ak} &\simeq&  - (p_1^+ + p_2^+) p_k^- \qquad k=1, 2 \nn\\ 
s_{bk} &\simeq& - p_k^+ p_3^- \qquad k=1, 2 \label{nmrinv}\\ 
s_{b3} &\simeq& - p_3^+ p_3^- = - p_{3\perp}^\ast p_{3\perp} \nn\\
s_{k3} &\simeq& p_k^+ p_3^- \qquad k=1, 2 \nonumber\\ 
s_{12} &=& p_1^+ p_2^- + p_1^- p_2^+ - p_{1\perp}^\ast p_{2\perp} - p_{2\perp}^\ast p_{1\perp} \,. \nonumber
\end{eqnarray}
The spinor products (\ref{spro}) become
\begin{eqnarray}
\langle p_a p_b\rangle &\simeq& - \sqrt{(p_1^+ + p_2^+) p_3^-}\,,\nonumber\\ 
\langle p_a p_3\rangle &=& -i \sqrt{-p_a^+\over p_3^+}\, p_{3\perp} \simeq i
{p_{3\perp}\over |p_{3\perp}|} \langle p_a p_b\rangle\, ,\nonumber\\
\langle p_a p_k\rangle &=& -i \sqrt{-p_a^+\over p_k^+}\, p_{k\perp}
\simeq -i \sqrt{p_1^+ + p_2^+\over p_k^+} p_{k\perp}\, ,
\nonumber \qquad k=1, 2\\
\langle p_k p_b\rangle &=& i \sqrt{-p_b^- p_k^+}\, 
\simeq i \sqrt{p_k^+ p_3^-}\, ,\label{frpro}\qquad k=1, 2\nonumber\\
\langle p_3 p_b\rangle &=& i \sqrt{-p_b^- p_3^+}\,
\simeq i |p_{3\perp}|\, ,\label{qmrkspro}\\
\langle p_k p_3\rangle &=& p_{k\perp}\sqrt{p_3^+\over p_k^+} - p_{3\perp}
\sqrt{p_k^+\over p_3^+} \simeq - p_{3\perp}\, \sqrt{p_k^+\over p_3^+}\, 
,\nonumber \qquad k=1, 2\\
\langle p_1 p_2\rangle &=& p_{1\perp}\sqrt{p_2^+\over p_1^+} - 
p_{2\perp}\sqrt{p_1^+\over p_2^+}\, .\nonumber
\end{eqnarray}

\section{Anomalous dimensions}
\label{AppAnDim}

The perturbative expansion of the
cusp anomalous dimension~\cite{Korchemsky:1985xj,Moch:2004pa}, divided by the relevant quadratic Casimir factor $C_i$, is
\beq
\label{hatgammaK}
  \gamma_K (\as)  = \sum_{L=1}^\infty \gamma_K^{(L)} \left( \frac{\alpha_s}{\pi} \right)^L\,,
\eeq
with 
\beq
\gamma_K^{(1)} = 2\,.
\label{eq:k2}
\eeq

The perturbative expansion of the collinear anomalous dimension is
\beq
\label{collad}
  \gamma_i (\as)  = \sum_{L=1}^\infty \gamma_i^{(L)} \left( \frac{\alpha_s}{\pi} \right)^L\,, \qquad i = q, g\,,
\eeq
with coefficients,
\beq
\gamma_g^{(1)} = - \frac{\beta_0}{4}\,, \qquad \gamma_q^{(1)} = - \frac{3}4 C_F\,,
\label{eq:c1}
\eeq
where $\beta_0$ is the coefficient of the beta function,
\beq
\beta_0 = \frac{11N_c - 2N_F}3\,,
\label{eq:b0k2}
\eeq
and
\beq
C_F = \frac{N_c^2-1}{2N_c}\,.
\eeq
Note that, as customary in the literature, the expansion in \Eqns{hatgammaK}{collad} is in $\alpha_s/\pi$,
while the impact factor (\ref{fullv}) and the Regge trajectory (\ref{alphb}) are expanded in $\alpha_s/4\pi$.

%
%

\section{One-loop five-gluon colour-ordered amplitudes}
\label{sec:1loopcolamp}

In this appendix, we report in the notation of Ref.~\cite{DelDuca:1998cx} the one-loop five-gluon colour-ordered amplitudes, which were computed in Ref.~\cite{Bern:1993mq}.
The  one-loop five-gluon colour-ordered amplitudes in \eqn{eq:oneLcol} can be written as~\cite{Bern:1993mq}
\begin{eqnarray}
m_{5:1}^{[1]} &=& \cg m_{5g}^{(0)} \left[ V^g + 4V^f + V^s + 4G^f + G^s \right] \nn\\
m_{5:1}^{[1/2]} &=& - \cg m_{5g}^{(0)} \left[ V^f + V^s + G^f + G^s \right] \, ,
\label{stri}
\end{eqnarray}
with $\cg$ as in \eqn{cgam}, and where the superscripts $g, f, s$ label contributions from
an $N=4$ supersymmetric multiplet, an $N=1$ chiral multiplet,
and a complex scalar, respectively. As done in Ref.~\cite{DelDuca:1998cx},
in \eqn{stri} we have factored out the tree five-gluon colour-ordered amplitude.

For the one-loop impact factor $A^{gg(1)}$ (\ref{eq:a1loop}), we need the one-loop five-gluon colour-ordered amplitudes only
in the helicity configurations which are nonzero at tree level.
We write the functions for the $(1,2,3,4,5)$ colour order for the two
relevant helicity configurations below.
The function obtained from the $N=4$ multiplet
is the same for both helicity configurations~\cite{Bern:1993mq}, 
\begin{equation}
V^g = -{1\over\epsilon^2} \sum_{j=1}^5
\left({\mu^2\over -s_{j,j+1}} \right)^{\epsilon} +
\sum_{j=1}^5 \log\left({-s_{j,j+1}\over -s_{j+1,j+2}}\right) \,
\log\left({-s_{j+2,j-2}\over -s_{j-2,j-1}}\right)
+ {5\over 6}\pi^2\, -{\delta_R\over 3}\, 
.\label{vertica}
\end{equation}
The other functions depend on the helicity configuration. We define
\begin{equation}
I_{ijkl} = \left[ ij \right] \langle jk\rangle \left[ kl \right]
\langle li\rangle\, .\label{comp}
\end{equation}
For the $(1^-,2^-,3^+,4^+,5^+)$ helicity configuration
we have~\cite{Bern:1993mq},
\begin{eqnarray}
V^f &=& -{5\over 2\epsilon} - {1\over 2} \left[\log\left({\mu^2\over -s_{23}}
\right) + \log\left({\mu^2\over -s_{51}} \right)\right] - 2\, \nonumber\\
G^f &=& {I_{1234}+I_{1245}\over 2\,s_{12}\, s_{51}}
L_0\left({-s_{23} \over -s_{51}}\right) \label{vertic}\\
G^s &=& -{G^f\over 3} + {I_{1234}\,I_{1245}\,(I_{1234}+I_{1245})
\over 3\,s_{12}^3\, s_{51}^3} L_2\left(-s_{23}\over
-s_{51}\right) \nonumber\\
&+&  {I_{1235}^2\over 3\,s_{12}^2 s_{23} s_{51} }\left(1-{s_{35}
\over s_{12}}\right) + {I_{1234}\,I_{1245}\over 6\,s_{12}^2\,s_{23}\,
s_{51}}\, ,\nonumber
\end{eqnarray}
while for the $(1^-,2^+,3^-,4^+,5^+)$ helicity configuration, 
we have
\begin{eqnarray}
V^f &=& -{5\over 2\epsilon} - {1\over 2} \left[\log\left({\mu^2\over -s_{34}}
\right) + \log\left({\mu^2\over -s_{51}} \right)\right] - 2\, \nonumber\\
G^f &=& - {I_{1325}+I_{1342}\over 2s_{13}\, s_{51}}
L_0\left({-s_{34}\over -s_{51}}\right) + {I_{1324}\, I_{1342}\over s_{13}^2\,
s_{51}^2} Ls_1\left({-s_{23}\over -s_{51}}, {-s_{34}\over
-s_{51}}\right) \nonumber\\ &+& {I_{1325}\, I_{1352}\over s_{13}^2\, 
s_{34}^2} Ls_1\left({-s_{12}\over -s_{34}}, 
{-s_{51}\over -s_{34}}\right) \label{vertid}\\
G^s &=& -{I_{1324}^2\, I_{1342}^2\over s_{13}^4\, s_{24}^2\, s_{51}^2}
\left[ 2 Ls_1\left({-s_{23}\over -s_{51}}, {-s_{34}\over
-s_{51}}\right) + L_1\left({-s_{23}\over -s_{51}}\right)
+ L_1\left({-s_{34}\over -s_{51}}\right)\right] \nonumber\\
&& - {I_{1325}^2\, I_{1352}^2\over s_{13}^4\, s_{25}^2\, s_{34}^2}
\left[2 Ls_1\left({-s_{12}\over -s_{34}}, {-s_{51}\over -s_{34}}\right)
+ L_1\left({-s_{12}\over -s_{34}}\right) + L_1\left({-s_{51}\over 
-s_{34}}\right) \right] \nonumber\\ &+&
{2\over 3} {I_{1324}^3\, I_{1342}\over s_{13}^4\, s_{24}\, s_{51}^3}
L_2\left(-s_{23}\over -s_{51}\right) +
{2\over 3} {I_{1352}^3\, I_{1325}\over s_{13}^4\, s_{25}\, s_{34}^3}
L_2\left(-s_{12}\over -s_{34}\right) \nonumber\\
&+& {1\over 3 s_{51}^3}\, L_2\left(-s_{34}\over -s_{51}\right)
\left[ - {I_{1325}\,I_{1342}\, (I_{1325}+I_{1342})\over s_{13}^3}
+ 2 {I_{1342}^3\, I_{1324}\over s_{13}^4\, s_{24}} +
2 {I_{1325}^3\, I_{1352}\over s_{13}^4\, s_{25}} \right] \nonumber\\
&+& {I_{1325}+I_{1342} \over 6\,s_{13}\, s_{51}} 
L_0\left({-s_{34} \over -s_{51}}\right) + {I_{1325}^2\, I_{1342}^2\,
\over 3\, s_{13}^4\, s_{23}\, s_{51}\, s_{34}\, s_{12} } \nonumber\\
&+& {I_{1324}^2\, I_{1342}^2 \over 3 \, s_{13}^4\,s_{23}\,s_{24}\,
s_{34}\, s_{51}} + {I_{1325}^2\, I_{1352}^2 \over 3\, 
s_{13}^4\,s_{25}\,s_{12}\, s_{34}\, s_{51}} 
- {I_{1342}\, I_{1325}\over 6\, s_{13}^2
s_{34}\, s_{51}}\, ,\nonumber
\end{eqnarray}
with 
\begin{eqnarray}
L_0(x) &=& {\log(x)\over 1-x}\nonumber\\
L_1(x) &=& {\log(x) +1-x\over (1-x)^2} \nonumber\\
L_2(x) &=& {1\over (1-x)^3} \left[\log(x) -{x\over 2} +{1\over 2x} \right] 
\label{logar}\\ Ls_1(x,y) &=& {1\over (1-x-y)^2}\, \left[ {\rm Li}_2(1-x) 
+ {\rm Li}_2(1-y) \right. \nonumber\\ && \left. + \log(x)\,\log(y) + 
(1-x-y) [ L_0(x)+L_0(y) ] - {\pi^2\over 6} \right]\, ,\nonumber
\end{eqnarray}
where ${\rm Li}_2$ is the dilogarithm.

For both the helicity configurations above, the functions $V^s$ and
$V^f$ are related by,
\begin{equation}
V^s = -{V^f\over 3} + {2\over 9}\, .\label{rel}
\end{equation}
In the expansion in $\epsilon$, eq.~(\ref{vertica})-(\ref{rel}) are 
valid to $O(\epsilon^0)$.
The colour-ordered amplitudes~(\ref{stri}) defined in terms of
eqs.~(\ref{vertica})-(\ref{rel}) are $\overline{\rm MS}$ regulated.

The colour-weighted sum,
\begin{equation}
m_{5:1} = N_c m_{5:1}^{[1]} + N_f m_{5:1}^{[1/2]} \,,
\label{eq:colweight}
\end{equation}
can be written as the sum of a universal
piece, which is the same for both helicity configurations, and a 
non-universal piece, which depends on the helicity configuration,
\begin{equation}
m_{5:1} = m_{5:1}^u + m_{5:1}^{nu}\ ,\label{decomp}
\end{equation}
where, using eqs.~(\ref{stri}), (\ref{vertica}) and (\ref{rel}), we can write
\begin{eqnarray} 
m_{5:1}^u &=& \cg\, m_{5g}^{(0)}\, N_c\, V^g\, ,\nn\\
m_{5:1}^{nu} &=& \cg\, m_{5g}^{(0)}\, \left[\beta_0 V^f + \left(
4N_c-N_f\right) G^f + \left(N_c-N_f\right) \left(G^s+{2\over
9}\right)\right]\, ,
\label{uni}
\end{eqnarray}
which is useful to single out the \msb ultraviolet counterterm.

\section{The one-loop five-gluon amplitude in NMRK}
\label{sec:1loopampnmrk}

We evaluate the colour-ordered amplitudes $m_{5:1}^{[1]}$ and $m_{5:1}^{[1/2]}$, \app{sec:1loopcolamp}, 
in the NMRK region of \eqn{qmrapp1} and \app{sec:appc}. We use the Mandelstam invariants (\ref{nmrinv})
and the prescription,
\beq
\log(-x) = \log(|x|) - i \pi \Theta(x)\,,
\eeq
in order to select the real part of the one-loop five-gluon amplitude. For the helicity configuration 
$(a^-,1^+,2^+,b^{'+},b^-)$, the real part of the $\overline{\rm MS}$-renormalised one-loop five-gluon amplitude is
\begin{equation}
\begin{split}
& \frac1{\cg\, m_{5g}^{(0)}} \RE\big[ m_{5;1}(p_a^-, p^+_1, p^+_2 ,p_{b'}^+,p_b^-) \big] \\\
&= N_c \left\{ -\frac{1}{\epsilon^2} \left[ 2 \left( \frac{\mu^2}{|p_{1\perp}|^2} \right)^\epsilon 
+ \left( \frac{\mu^2}{|p_{2\perp}|^2} \right)^\epsilon + 2 \left( \frac{\mu^2}{-t} \right)^\epsilon \right] \right. \\\ 
&\qquad\quad + \frac{2}{\epsilon} \left[\left(\frac{\mu^2}{|p_{1\perp}|^2}\right)^\epsilon \log\left(\frac{s_{12}}{|p_{1\perp}| \: |p_{2\perp}|}\right)
+ \left(\frac{\mu^2}{-t}\right)^\epsilon \log\left(\frac{x_2\, s}{\sqrt{-t} \, |p_{2\perp}|}\right) \right. \\\
& \qquad\qquad \left. -\frac{1}{2} \left( \frac{\mu^2}{-t}\frac{|p_{2\perp}|^2}{|p_{1\perp}|^2} \left( \frac{p_2^+}{p_1^+}\frac{s_{12}}{|p_{2\perp}|^2} \right)^{\frac{1}{2}} \right)^\epsilon  \log \left( \frac{x_1x_2s_{12}}{|p_{2\perp}|^2} \right) \right]  \\\
& \left. \qquad\quad   -\frac{1}{2} \log^2\left(\frac{|p_{1\perp}|^2}{-t}\right) 
- \frac{1}{2}\log\left(x_1 \right)\log\left(x_1x_2\left(\frac{-t}{s_{12}}\right)^3 \frac{-t}{|p_{2\perp}|^2} \right) 
- \frac{64}{9} - \frac{\delta_R}{3} + \frac{4}{3}\pi^2\right\} \\\
& - \frac{\beta_0}{2\epsilon} \left[\left(\frac{x_1 \mu^2}{|p_{1\perp}|^2 }\right)^\epsilon 
+ \left(\frac{\mu^2}{-t} \right)^\epsilon \right]
+ \frac{10}{9}N_f + \frac{\beta_0}{2} \left(\frac{|p_{1\perp}|^2}{x_1} - t + 2q_\perp^*p_{1\perp}\right) \frac{L_0\left(\frac{|p_{1\perp}|^2}{-x_1t}\right)}{t} \\\
&+ \frac{N_C-N_f}{3} \left\{ \frac{p_2^+}{p_1^+}\left[ |p_{1\perp}|^2 \: {q_\perp^*}^2 \: p_{2\perp}^2
-2\: |p_{1\perp}|^2 \: |p_{2\perp}|^2 p_{1\perp} q_\perp^*  + \frac{p_2^+}{p_1^+} |p_{1\perp}|^4 \, p_{2\perp} q_\perp^*\right] \right. \\\
& \qquad\qquad\qquad  - |p_{2\perp}|^2  \left[2 \: |p_{1\perp}|^2(-t)+\left( 2\,|p_{2\perp}|^2 - |p_{1\perp}|^2 + t\right) p_{1\perp} q_\perp^*\right] \bigg\}
\frac{L_2\left(\frac{|p_{1\perp}|^2}{-x_1t}\right)}{t^3} \\\
&+\frac{N_C-N_f}{6} \left[\frac{x_2 p_{2\perp} q_\perp^*}{ -t} + \frac{2x_1x_2 {q_\perp^*}^2 p_{1\perp}^2}{- t\, |p_{1\perp}|^2} 
- \frac{ x_1 |p_{2\perp}|^2 p_{1\perp} q_\perp^*}{ -t\, |p_{1\perp}|^2} \right] 
- \frac{3}{2} \frac{\beta_0}{\epsilon}\,,
\end{split}
\label{oneloop1}
\end{equation}
where the last term is the $\overline{\rm MS}$ ultraviolet counterterm,
where $q_\perp = -(p_{1\perp} +p_{2\perp} )$, with $t= - |q_\perp|^2$, and where it is understood that 
we must use
\beqa
2\, \RE (q_\perp^*p_{1\perp} ) &=& t + |p_{2\perp}|^2 - |p_{1\perp}|^2\,, \nn\\
2\, \RE (q_\perp^*p_{2\perp} ) &=& t + |p_{1\perp}|^2 - |p_{2\perp}|^2 \,.
\label{eq:realpart}
\eeqa

In the MRK, $p_1^+\gg p_2^+$, the quantities that follow become,
\beqa
x_1 &\rightarrow& 1 \nn\\
x_2 &\rightarrow& \frac{p_2^+}{p_1^+} \nn\\
s_{12} &\rightarrow& \frac{p_1^+}{p_2^+} |p_{2\perp}|^2 \nn\\
 \log\left(\frac{s_{12}}{|p_{1\perp}| \: |p_{2\perp}|}\right) &\rightarrow& \log\left(\frac{s_{12}}{\sqrt{-t_1} \: |p_{2\perp}|}\right) \nn\\
 \log\left(\frac{x_2\, s}{\sqrt{-t} \, |p_{2\perp}|}\right) &\rightarrow& \log\left(\frac{s_{23}}{\sqrt{-t_2} \: |p_{2\perp}|}\right)\,,
 \label{eq:x1x2mrk}
\eeqa
with $t_1 = - |p_{1\perp}|^2$ and $t_2 = - |p_{b'\perp}|^2$.
In particular, the logarithms in the last two lines of \eqn{eq:x1x2mrk}, which occur on the second line of the right-hand side of \eqn{oneloop1} 
become the two rapidity intervals of a five-gluon amplitude in MRK,
and \eqn{oneloop1} is reduced to the one-loop five-gluon amplitude in MRK~\cite{DelDuca:1998cx}.

\section{Coefficients of the one-loop two-quark three-gluon amplitude in NMRK}
\label{sec:g}

Here we display the real parts of the coefficients $\lambda_1$ and $\chi_1$, which are factored out in \eqns{eq:a12qq}{eq:1/2bba12},
\begin{align}
\frac1{\cg} \RE\big[ \lambda_1 \big] = &  -\frac{1}{\epsilon^2} \left[ 2 \left( \frac{\mu^2}{|p_{1\perp}|^2} \right)^{\epsilon} +  \left( \frac{\mu^2}{|p_{2\perp}|^2} \right)^{\epsilon} + \left( \frac{\mu^2}{-t} \right)^\epsilon \right] \nn\\
&+ \frac{2}{\epsilon} \left[ \left(\frac{\mu^2}{|p_{1\perp}|^2}\right)^\epsilon \log\left(\frac{s_{12}}{|p_{1\perp}| \: |p_{2\perp}|}\right) 
+ \left( \frac{\mu^2}{-t} \right)^{\epsilon} \log\left(\frac{x_2\, s}{\sqrt{-t} \, |p_{2\perp}|}\right) \right. \nn\\
& \left. -\frac{1}{2} \left( \frac{\mu^2}{-t}\frac{|p_{2\perp}|^2}{|p_{1\perp}|^2} \left( \frac{p_2^+}{p_1^+}\frac{s_{12}}{|p_{2\perp}|^2} \right)^{\frac{1}{2}} \right)^\epsilon  \log \left( \frac{x_1x_2s_{12}}{|p_{2\perp}|^2} \right) \right] \nn\\
&-  \left( \frac{\mu^2}{-t} \right)^\epsilon \left( \frac{3}{2\epsilon} - \frac{83}{18}  \right) 
-  \left( \frac{\mu^2}{|p_{1\perp}|^2} \right)^\epsilon \frac{11}{3 \epsilon} - \frac{11}{2 \epsilon}  \nn\\
&- \frac{1}{2} \log^2\left(\frac{|p_{1\perp}|^2}{-t}\right) 
- \frac{1}{2}\log\left(x_1 \right)\log\left(x_1x_2\left(\frac{-t}{s_{12}}\right)^3 \frac{-t}{|p_{2\perp}|^2} \right) \nn\\
&- \frac{11}{6}\log \left( x_1 \right) + \frac{4}{3}\pi^2   -\frac{64}{9} 
+ \frac{11}{6} \left(\frac{|p_{1\perp}|^2}{x_1} - t + 2q_\perp^*p_{1\perp}\right) \frac{L_0\left(\frac{|p_{1\perp}|^2}{-x_1t}\right)}{t} \nn\\
&+ \frac1{3} \left\{ \frac{p_2^+}{p_1^+}\left[ |p_{1\perp}|^2 \: {q_\perp^*}^2 \: p_{2\perp}^2
-2\: |p_{1\perp}|^2 \: |p_{2\perp}|^2 p_{1\perp} q_\perp^*  + \frac{p_2^+}{p_1^+} |p_{1\perp}|^4 \, p_{2\perp} q_\perp^*\right] \right. \nn\\
& \qquad  - |p_{2\perp}|^2  \left[2 \: |p_{1\perp}|^2(-t)+\left( 2\,|p_{2\perp}|^2 - |p_{1\perp}|^2 + t\right) p_{1\perp} q_\perp^*\right] \bigg\}
\frac{L_2\left(\frac{|p_{1\perp}|^2}{-x_1t}\right)}{t^3} \nn\\
&+ \frac{1}{6} \left[\frac{x_2 p_{2\perp} q_\perp^*}{ -t} + \frac{2x_1x_2 {q_\perp^*}^2 p_{1\perp}^2}{- t\, |p_{1\perp}|^2} 
- \frac{ x_1 |p_{2\perp}|^2 p_{1\perp} q_\perp^*}{ -t\, |p_{1\perp}|^2} \right] \,,
\label{eq:g1} 
\end{align}
and $\lambda_2$ is obtained from $\lambda_1$ by swapping the labels 1 and 2,
\beq
\lambda_2 = \lambda_1(1\leftrightarrow 2)\,.
\label{eq:g2} 
\eeq
\begin{align}
\frac1{\cg} \RE\big[ \chi_1 \big] &=  \frac{1}{\epsilon} + \frac{2}{3 \epsilon} \left( \frac{\mu^2}{|p_{1\perp}|^2} \right)^\epsilon - \frac{10}{9} \left( \frac{\mu^2}{-t} \right)^\epsilon + \frac{1}{3} \log \left( x_1 \right) + \frac{10}{9} \nn \\
&- \frac{1}{3} \left(\frac{|p_{1\perp}|^2}{x_1} - t + 2q_\perp^*p_{1\perp}\right) \frac{L_0\left(\frac{|p_{1\perp}|^2}{-x_1t}\right)}{t} \nn\\
&- \frac{1}{3} \left\{ \frac{p_2^+}{p_1^+}\left[ |p_{1\perp}|^2 \: {q_\perp^*}^2 \: p_{2\perp}^2
-2\: |p_{1\perp}|^2 \: |p_{2\perp}|^2 p_{1\perp} q_\perp^*  + \frac{p_2^+}{p_1^+} |p_{1\perp}|^4 \, p_{2\perp} q_\perp^*\right] \right. \nn\\
& \qquad  - |p_{2\perp}|^2  \left[2 \: |p_{1\perp}|^2(-t)+\left( 2\,|p_{2\perp}|^2 - |p_{1\perp}|^2 + t\right) p_{1\perp} q_\perp^*\right] \bigg\}
\frac{L_2\left(\frac{|p_{1\perp}|^2}{-x_1t}\right)}{t^3} \nn\\
&-  \frac{1}{6} \left[\frac{x_2 p_{2\perp} q_\perp^*}{ -t} + \frac{2x_1x_2 {q_\perp^*}^2 p_{1\perp}^2}{- t\, |p_{1\perp}|^2} 
- \frac{ x_1 |p_{2\perp}|^2 p_{1\perp} q_\perp^*}{ -t\, |p_{1\perp}|^2} \right] \,,
\label{eq:g3} 
\end{align}
and $\chi_2$ is obtained from $\chi_1$ by swapping the labels 1 and 2,
\beq
\chi_2 = \chi_1(1\leftrightarrow 2)\,,
\label{eq:g4} 
\eeq
where it is understood that in order to obtain the real part of \eqnss{eq:g1}{eq:g4}, we must use \eqn{eq:realpart}.

\bibliography{refs}

\providecommand{\href}[2]{#2}\begingroup\raggedright\begin{thebibliography}{10}

\bibitem{Lipatov:1976zz}
L.~N. Lipatov, {\it {Reggeization of the Vector Meson and the Vacuum
  Singularity in Nonabelian Gauge Theories}},  {\em Sov. J. Nucl. Phys.} {\bf
  23} (1976) 338--345. [Yad. Fiz.23,642(1976)].

\bibitem{Dixon:2012yy}
L.~J. Dixon, C.~Duhr, and J.~Pennington, {\it {Single-valued harmonic
  polylogarithms and the multi-Regge limit}},  {\em JHEP} {\bf 10} (2012) 074,
  [\href{http://xxx.lanl.gov/abs/1207.0186}{{\tt 1207.0186}}].

\bibitem{DelDuca:2013lma}
V.~Del~Duca, L.~J. Dixon, C.~Duhr, and J.~Pennington, {\it {The BFKL equation,
  Mueller-Navelet jets and single-valued harmonic polylogarithms}},  {\em JHEP}
  {\bf 02} (2014) 086, [\href{http://xxx.lanl.gov/abs/1309.6647}{{\tt
  1309.6647}}].

\bibitem{DelDuca:2016lad}
V.~Del~Duca, S.~Druc, J.~Drummond, C.~Duhr, F.~Dulat, R.~Marzucca,
  G.~Papathanasiou, and B.~Verbeek, {\it {Multi-Regge kinematics and the moduli
  space of Riemann spheres with marked points}},  {\em JHEP} {\bf 08} (2016)
  152, [\href{http://xxx.lanl.gov/abs/1606.08807}{{\tt 1606.08807}}].

\bibitem{DelDuca:2017peo}
V.~Del~Duca, C.~Duhr, R.~Marzucca, and B.~Verbeek, {\it {The analytic structure
  and the transcendental weight of the BFKL ladder at NLL accuracy}},  {\em
  JHEP} {\bf 10} (2017) 001, [\href{http://xxx.lanl.gov/abs/1705.10163}{{\tt
  1705.10163}}].

\bibitem{DelDuca:2019tur}
V.~Del~Duca, S.~Druc, J.~M. Drummond, C.~Duhr, F.~Dulat, R.~Marzucca,
  G.~Papathanasiou, and B.~Verbeek, {\it {All-order amplitudes at any
  multiplicity in the multi-Regge limit}},  {\em Phys. Rev. Lett.} {\bf 124}
  (2020), no.~16 161602, [\href{http://xxx.lanl.gov/abs/1912.00188}{{\tt
  1912.00188}}].

\bibitem{Caron-Huot:2020grv}
S.~Caron-Huot, E.~Gardi, J.~Reichel, and L.~Vernazza, {\it {Two-parton
  scattering amplitudes in the Regge limit to high loop orders}},  {\em JHEP}
  {\bf 08} (2020) 116, [\href{http://xxx.lanl.gov/abs/2006.01267}{{\tt
  2006.01267}}].

\bibitem{Kuraev:1976ge}
E.~A. Kuraev, L.~N. Lipatov, and V.~S. Fadin, {\it {Multi - Reggeon Processes
  in the Yang-Mills Theory}},  {\em Sov. Phys. JETP} {\bf 44} (1976) 443--450.
  [Zh. Eksp. Teor. Fiz.71,840(1976)].

\bibitem{Bern:2008qj}
Z.~Bern, J.~J.~M. Carrasco, and H.~Johansson, {\it {New Relations for
  Gauge-Theory Amplitudes}},  {\em Phys. Rev. D} {\bf 78} (2008) 085011,
  [\href{http://xxx.lanl.gov/abs/0805.3993}{{\tt 0805.3993}}].

\bibitem{DelDuca:1995zy}
V.~Del~Duca, {\it {Equivalence of the Parke-Taylor and the Fadin-Kuraev-Lipatov
  amplitudes in the high-energy limit}},  {\em Phys. Rev. D} {\bf 52} (1995)
  1527--1534, [\href{http://xxx.lanl.gov/abs/hep-ph/9503340}{{\tt
  hep-ph/9503340}}].

\bibitem{Balitsky:1979ap}
I.~I. Balitsky, L.~N. Lipatov, and V.~S. Fadin, {\it {REGGE PROCESSES IN
  NONABELIAN GAUGE THEORIES. (IN RUSSIAN)}}, .

\bibitem{Fadin:1975cb}
V.~S. Fadin, E.~A. Kuraev, and L.~N. Lipatov, {\it {On the Pomeranchuk
  Singularity in Asymptotically Free Theories}},  {\em Phys. Lett.} {\bf B60}
  (1975) 50--52.

\bibitem{Kuraev:1977fs}
E.~A. Kuraev, L.~N. Lipatov, and V.~S. Fadin, {\it {The Pomeranchuk Singularity
  in Nonabelian Gauge Theories}},  {\em Sov. Phys. JETP} {\bf 45} (1977)
  199--204. [Zh. Eksp. Teor. Fiz.72,377(1977)].

\bibitem{Balitsky:1978ic}
I.~I. Balitsky and L.~N. Lipatov, {\it {The Pomeranchuk Singularity in Quantum
  Chromodynamics}},  {\em Sov. J. Nucl. Phys.} {\bf 28} (1978) 822--829. [Yad.
  Fiz.28,1597(1978)].

\bibitem{Mueller:1986ey}
A.~H. Mueller and H.~Navelet, {\it {An Inclusive Minijet Cross-Section and the
  Bare Pomeron in QCD}},  {\em Nucl. Phys. B} {\bf 282} (1987) 727--744.

\bibitem{DelDuca:1993mn}
V.~Del~Duca and C.~R. Schmidt, {\it {Dijet production at large rapidity
  intervals}},  {\em Phys. Rev. D} {\bf 49} (1994) 4510--4516,
  [\href{http://xxx.lanl.gov/abs/hep-ph/9311290}{{\tt hep-ph/9311290}}].

\bibitem{Stirling:1994he}
W.~J. Stirling, {\it {Production of jet pairs at large relative rapidity in
  hadron hadron collisions as a probe of the perturbative pomeron}},  {\em
  Nucl. Phys. B} {\bf 423} (1994) 56--79,
  [\href{http://xxx.lanl.gov/abs/hep-ph/9401266}{{\tt hep-ph/9401266}}].

\bibitem{Andersen:2001kta}
J.~R. Andersen, V.~Del~Duca, S.~Frixione, C.~R. Schmidt, and W.~J. Stirling,
  {\it {Mueller-Navelet jets at hadron colliders}},  {\em JHEP} {\bf 02} (2001)
  007, [\href{http://xxx.lanl.gov/abs/hep-ph/0101180}{{\tt hep-ph/0101180}}].

\bibitem{Andersen:2008gc}
J.~R. Andersen, V.~Del~Duca, and C.~D. White, {\it {Higgs Boson Production in
  Association with Multiple Hard Jets}},  {\em JHEP} {\bf 02} (2009) 015,
  [\href{http://xxx.lanl.gov/abs/0808.3696}{{\tt 0808.3696}}].

\bibitem{Andersen:2009nu}
J.~R. Andersen and J.~M. Smillie, {\it {Constructing All-Order Corrections to
  Multi-Jet Rates}},  {\em JHEP} {\bf 01} (2010) 039,
  [\href{http://xxx.lanl.gov/abs/0908.2786}{{\tt 0908.2786}}].

\bibitem{Andersen:2011hs}
J.~R. Andersen and J.~M. Smillie, {\it {Multiple Jets at the LHC with High
  Energy Jets}},  {\em JHEP} {\bf 06} (2011) 010,
  [\href{http://xxx.lanl.gov/abs/1101.5394}{{\tt 1101.5394}}].

\bibitem{Fadin:1993wh}
V.~S. Fadin and L.~N. Lipatov, {\it {Radiative corrections to QCD scattering
  amplitudes in a multi - Regge kinematics}},  {\em Nucl. Phys. B} {\bf 406}
  (1993) 259--292.

\bibitem{Fadin:1995xg}
V.~S. Fadin, M.~I. Kotsky, and R.~Fiore, {\it {Gluon Reggeization in QCD in the
  next-to-leading order}},  {\em Phys. Lett. B} {\bf 359} (1995) 181--188.

\bibitem{Fadin:1996tb}
V.~S. Fadin, R.~Fiore, and M.~I. Kotsky, {\it {Gluon Regge trajectory in the
  two loop approximation}},  {\em Phys. Lett. B} {\bf 387} (1996) 593--602,
  [\href{http://xxx.lanl.gov/abs/hep-ph/9605357}{{\tt hep-ph/9605357}}].

\bibitem{Fadin:1995km}
V.~S. Fadin, R.~Fiore, and A.~Quartarolo, {\it {Reggeization of quark quark
  scattering amplitude in QCD}},  {\em Phys. Rev. D} {\bf 53} (1996)
  2729--2741, [\href{http://xxx.lanl.gov/abs/hep-ph/9506432}{{\tt
  hep-ph/9506432}}].

\bibitem{Blumlein:1998ib}
J.~Blumlein, V.~Ravindran, and W.~L. van Neerven, {\it {On the gluon Regge
  trajectory in O alpha-s**2}},  {\em Phys. Rev. D} {\bf 58} (1998) 091502,
  [\href{http://xxx.lanl.gov/abs/hep-ph/9806357}{{\tt hep-ph/9806357}}].

\bibitem{DelDuca:2001gu}
V.~Del~Duca and E.~W.~N. Glover, {\it {The High-energy limit of QCD at two
  loops}},  {\em JHEP} {\bf 10} (2001) 035,
  [\href{http://xxx.lanl.gov/abs/hep-ph/0109028}{{\tt hep-ph/0109028}}].

\bibitem{Fadin:1992zt}
V.~S. Fadin and R.~Fiore, {\it {Quark contribution to the gluon-gluon - reggeon
  vertex in QCD}},  {\em Phys. Lett. B} {\bf 294} (1992) 286--292.

\bibitem{Fadin:1993qb}
V.~S. Fadin, R.~Fiore, and A.~Quartarolo, {\it {Radiative corrections to quark
  quark reggeon vertex in QCD}},  {\em Phys. Rev. D} {\bf 50} (1994)
  2265--2276, [\href{http://xxx.lanl.gov/abs/hep-ph/9310252}{{\tt
  hep-ph/9310252}}].

\bibitem{DelDuca:1998kx}
V.~Del~Duca and C.~R. Schmidt, {\it {Virtual next-to-leading corrections to the
  impact factors in the high-energy limit}},  {\em Phys. Rev. D} {\bf 57}
  (1998) 4069--4079, [\href{http://xxx.lanl.gov/abs/hep-ph/9711309}{{\tt
  hep-ph/9711309}}].

\bibitem{Bern:1998sc}
Z.~Bern, V.~Del~Duca, and C.~R. Schmidt, {\it {The Infrared behavior of one
  loop gluon amplitudes at next-to-next-to-leading order}},  {\em Phys. Lett.
  B} {\bf 445} (1998) 168--177,
  [\href{http://xxx.lanl.gov/abs/hep-ph/9810409}{{\tt hep-ph/9810409}}].

\bibitem{Fadin:2006bj}
V.~S. Fadin, R.~Fiore, M.~G. Kozlov, and A.~V. Reznichenko, {\it {Proof of the
  multi-Regge form of QCD amplitudes with gluon exchanges in the NLA}},  {\em
  Phys. Lett.} {\bf B639} (2006) 74--81,
  [\href{http://xxx.lanl.gov/abs/hep-ph/0602006}{{\tt hep-ph/0602006}}].

\bibitem{Fadin:2015zea}
V.~S. Fadin, M.~G. Kozlov, and A.~V. Reznichenko, {\it {Gluon Reggeization in
  Yang-Mills Theories}},  {\em Phys. Rev.} {\bf D92} (2015), no.~8 085044,
  [\href{http://xxx.lanl.gov/abs/1507.00823}{{\tt 1507.00823}}].

\bibitem{Fadin:1998py}
V.~S. Fadin and L.~N. Lipatov, {\it {BFKL pomeron in the next-to-leading
  approximation}},  {\em Phys. Lett.} {\bf B429} (1998) 127--134,
  [\href{http://xxx.lanl.gov/abs/hep-ph/9802290}{{\tt hep-ph/9802290}}].

\bibitem{Ciafaloni:1998gs}
M.~Ciafaloni and G.~Camici, {\it {Energy scale(s) and next-to-leading BFKL
  equation}},  {\em Phys. Lett.} {\bf B430} (1998) 349--354,
  [\href{http://xxx.lanl.gov/abs/hep-ph/9803389}{{\tt hep-ph/9803389}}].

\bibitem{Fadin:1989kf}
V.~S. Fadin and L.~N. Lipatov, {\it {High-Energy Production of Gluons in a
  QuasimultiRegge Kinematics}},  {\em JETP Lett.} {\bf 49} (1989) 352.

\bibitem{DelDuca:1995ki}
V.~Del~Duca, {\it {Real next-to-leading corrections to the multi - gluon
  amplitudes in the helicity formalism}},  {\em Phys. Rev. D} {\bf 54} (1996)
  989--1009, [\href{http://xxx.lanl.gov/abs/hep-ph/9601211}{{\tt
  hep-ph/9601211}}].

\bibitem{Fadin:1996nw}
V.~S. Fadin and L.~N. Lipatov, {\it {Next-to-leading corrections to the BFKL
  equation from the gluon and quark production}},  {\em Nucl. Phys. B} {\bf
  477} (1996) 767--808, [\href{http://xxx.lanl.gov/abs/hep-ph/9602287}{{\tt
  hep-ph/9602287}}].

\bibitem{DelDuca:1996nom}
V.~Del~Duca, {\it {Quark - anti-quark contribution to the multi - gluon
  amplitudes in the helicity formalism}},  {\em Phys. Rev. D} {\bf 54} (1996)
  4474--4482, [\href{http://xxx.lanl.gov/abs/hep-ph/9604250}{{\tt
  hep-ph/9604250}}].

\bibitem{DelDuca:1996km}
V.~Del~Duca, {\it {Next-to-leading corrections to the BFKL equation}},  {\em
  Frascati Phys. Ser.} {\bf 5} (1996) 463--478,
  [\href{http://xxx.lanl.gov/abs/hep-ph/9605404}{{\tt hep-ph/9605404}}].

\bibitem{Fadin:1994fj}
V.~S. Fadin, R.~Fiore, and A.~Quartarolo, {\it {Quark contribution to the
  reggeon - reggeon - gluon vertex in QCD}},  {\em Phys. Rev. D} {\bf 50}
  (1994) 5893--5901, [\href{http://xxx.lanl.gov/abs/hep-th/9405127}{{\tt
  hep-th/9405127}}].

\bibitem{Fadin:1996yv}
V.~S. Fadin, R.~Fiore, and M.~I. Kotsky, {\it {Gribov's theorem on soft
  emission and the reggeon-reggeon - gluon vertex at small transverse
  momentum}},  {\em Phys. Lett. B} {\bf 389} (1996) 737--741,
  [\href{http://xxx.lanl.gov/abs/hep-ph/9608229}{{\tt hep-ph/9608229}}].

\bibitem{DelDuca:1998cx}
V.~Del~Duca and C.~R. Schmidt, {\it {Virtual next-to-leading corrections to the
  Lipatov vertex}},  {\em Phys. Rev. D} {\bf 59} (1999) 074004,
  [\href{http://xxx.lanl.gov/abs/hep-ph/9810215}{{\tt hep-ph/9810215}}].

\bibitem{Colferai:2010wu}
D.~Colferai, F.~Schwennsen, L.~Szymanowski, and S.~Wallon, {\it {Mueller
  Navelet jets at LHC - complete NLL BFKL calculation}},  {\em JHEP} {\bf 12}
  (2010) 026, [\href{http://xxx.lanl.gov/abs/1002.1365}{{\tt 1002.1365}}].

\bibitem{Ducloue:2013hia}
B.~Ducloue, L.~Szymanowski, and S.~Wallon, {\it {Confronting Mueller-Navelet
  jets in NLL BFKL with LHC experiments at 7 TeV}},  {\em JHEP} {\bf 05} (2013)
  096, [\href{http://xxx.lanl.gov/abs/1302.7012}{{\tt 1302.7012}}].

\bibitem{Ducloue:2013bva}
B.~Duclou\'e, L.~Szymanowski, and S.~Wallon, {\it {Evidence for high-energy
  resummation effects in Mueller-Navelet jets at the LHC}},  {\em Phys. Rev.
  Lett.} {\bf 112} (2014) 082003,
  [\href{http://xxx.lanl.gov/abs/1309.3229}{{\tt 1309.3229}}].

\bibitem{Bartels:2001ge}
J.~Bartels, D.~Colferai, and G.~P. Vacca, {\it {The NLO jet vertex for
  Mueller-Navelet and forward jets: The Quark part}},  {\em Eur. Phys. J. C}
  {\bf 24} (2002) 83--99, [\href{http://xxx.lanl.gov/abs/hep-ph/0112283}{{\tt
  hep-ph/0112283}}].

\bibitem{Bartels:2002yj}
J.~Bartels, D.~Colferai, and G.~P. Vacca, {\it {The NLO jet vertex for
  Mueller-Navelet and forward jets: The Gluon part}},  {\em Eur. Phys. J. C}
  {\bf 29} (2003) 235--249, [\href{http://xxx.lanl.gov/abs/hep-ph/0206290}{{\tt
  hep-ph/0206290}}].

\bibitem{Caron-Huot:2017fxr}
S.~Caron-Huot, E.~Gardi, and L.~Vernazza, {\it {Two-parton scattering in the
  high-energy limit}},  {\em JHEP} {\bf 06} (2017) 016,
  [\href{http://xxx.lanl.gov/abs/1701.05241}{{\tt 1701.05241}}].

\bibitem{Caron-Huot:2013fea}
S.~Caron-Huot, {\it {When does the gluon reggeize?}},  {\em JHEP} {\bf 05}
  (2015) 093, [\href{http://xxx.lanl.gov/abs/1309.6521}{{\tt 1309.6521}}].

\bibitem{Caron-Huot:2017zfo}
S.~Caron-Huot, E.~Gardi, J.~Reichel, and L.~Vernazza, {\it {Infrared
  singularities of QCD scattering amplitudes in the Regge limit to all
  orders}},  {\em JHEP} {\bf 03} (2018) 098,
  [\href{http://xxx.lanl.gov/abs/1711.04850}{{\tt 1711.04850}}].

\bibitem{Fadin:2016wso}
V.~S. Fadin, {\it {Particularities of the NNLLA BFKL}},  {\em AIP Conf. Proc.}
  {\bf 1819} (2017), no.~1 060003,
  [\href{http://xxx.lanl.gov/abs/1612.04481}{{\tt 1612.04481}}].

\bibitem{Fadin:2017nka}
V.~S. Fadin and L.~N. Lipatov, {\it {Reggeon cuts in QCD amplitudes with
  negative signature}},  {\em Eur. Phys. J. C} {\bf 78} (2018), no.~6 439,
  [\href{http://xxx.lanl.gov/abs/1712.09805}{{\tt 1712.09805}}].

\bibitem{Falcioni:2020lvv}
G.~Falcioni, E.~Gardi, C.~Milloy, and L.~Vernazza, {\it {Climbing three-Reggeon
  ladders: four-loop amplitudes in the high-energy limit in full colour}},
  \href{http://xxx.lanl.gov/abs/2012.00613}{{\tt 2012.00613}}.

\bibitem{Bret:2011xm}
V.~Del~Duca, C.~Duhr, E.~Gardi, L.~Magnea, and C.~D. White, {\it {An infrared
  approach to Reggeization}},  {\em Phys. Rev. D} {\bf 85} (2012) 071104,
  [\href{http://xxx.lanl.gov/abs/1108.5947}{{\tt 1108.5947}}].

\bibitem{DelDuca:2011ae}
V.~Del~Duca, C.~Duhr, E.~Gardi, L.~Magnea, and C.~D. White, {\it {The Infrared
  structure of gauge theory amplitudes in the high-energy limit}},  {\em JHEP}
  {\bf 12} (2011) 021, [\href{http://xxx.lanl.gov/abs/1109.3581}{{\tt
  1109.3581}}].

\bibitem{DelDuca:2013ara}
V.~Del~Duca, G.~Falcioni, L.~Magnea, and L.~Vernazza, {\it {High-energy QCD
  amplitudes at two loops and beyond}},  {\em Phys. Lett. B} {\bf 732} (2014)
  233--240, [\href{http://xxx.lanl.gov/abs/1311.0304}{{\tt 1311.0304}}].

\bibitem{DelDuca:2014cya}
V.~Del~Duca, G.~Falcioni, L.~Magnea, and L.~Vernazza, {\it {Analyzing
  high-energy factorization beyond next-to-leading logarithmic accuracy}},
  {\em JHEP} {\bf 02} (2015) 029,
  [\href{http://xxx.lanl.gov/abs/1409.8330}{{\tt 1409.8330}}].

\bibitem{Korchemskaya:1996je}
I.~A. Korchemskaya and G.~P. Korchemsky, {\it {Evolution equation for gluon
  Regge trajectory}},  {\em Phys. Lett. B} {\bf 387} (1996) 346--354,
  [\href{http://xxx.lanl.gov/abs/hep-ph/9607229}{{\tt hep-ph/9607229}}].

\bibitem{Henn:2016jdu}
J.~M. Henn and B.~Mistlberger, {\it {Four-Gluon Scattering at Three Loops,
  Infrared Structure, and the Regge Limit}},  {\em Phys. Rev. Lett.} {\bf 117}
  (2016), no.~17 171601, [\href{http://xxx.lanl.gov/abs/1608.00850}{{\tt
  1608.00850}}].

\bibitem{Lipatov:2009nt}
L.~N. Lipatov, {\it {Integrability of scattering amplitudes in N=4 SUSY}},
  {\em J. Phys. A} {\bf 42} (2009) 304020,
  [\href{http://xxx.lanl.gov/abs/0902.1444}{{\tt 0902.1444}}].

\bibitem{DelDuca:1999iql}
V.~Del~Duca, A.~Frizzo, and F.~Maltoni, {\it {Factorization of tree QCD
  amplitudes in the high-energy limit and in the collinear limit}},  {\em Nucl.
  Phys. B} {\bf 568} (2000) 211--262,
  [\href{http://xxx.lanl.gov/abs/hep-ph/9909464}{{\tt hep-ph/9909464}}].

\bibitem{Antonov:2004hh}
E.~N. Antonov, L.~N. Lipatov, E.~A. Kuraev, and I.~O. Cherednikov, {\it
  {Feynman rules for effective Regge action}},  {\em Nucl. Phys. B} {\bf 721}
  (2005) 111--135, [\href{http://xxx.lanl.gov/abs/hep-ph/0411185}{{\tt
  hep-ph/0411185}}].

\bibitem{Duhr:2009uxa}
C.~Duhr, {\em {New techniques in QCD}}.
\newblock PhD thesis, Louvain U., CP3, 2009.

\bibitem{Bern:1993mq}
Z.~Bern, L.~J. Dixon, and D.~A. Kosower, {\it {One loop corrections to five
  gluon amplitudes}},  {\em Phys. Rev. Lett.} {\bf 70} (1993) 2677--2680,
  [\href{http://xxx.lanl.gov/abs/hep-ph/9302280}{{\tt hep-ph/9302280}}].

\bibitem{Bern:1994fz}
Z.~Bern, L.~J. Dixon, and D.~A. Kosower, {\it {One loop corrections to two
  quark three gluon amplitudes}},  {\em Nucl. Phys. B} {\bf 437} (1995)
  259--304, [\href{http://xxx.lanl.gov/abs/hep-ph/9409393}{{\tt
  hep-ph/9409393}}].

\bibitem{DelDuca:2017pmn}
V.~Del~Duca, {\it {Iterating QCD scattering amplitudes in the high-energy
  limit}},  {\em JHEP} {\bf 02} (2018) 112,
  [\href{http://xxx.lanl.gov/abs/1712.07030}{{\tt 1712.07030}}].

\bibitem{Catani:1996pk}
S.~Catani, M.~H. Seymour, and Z.~Trocsanyi, {\it {Regularization scheme
  independence and unitarity in QCD cross-sections}},  {\em Phys. Rev. D} {\bf
  55} (1997) 6819--6829, [\href{http://xxx.lanl.gov/abs/hep-ph/9610553}{{\tt
  hep-ph/9610553}}].

\bibitem{DelDuca:1999rs}
V.~Del~Duca, L.~J. Dixon, and F.~Maltoni, {\it {New color decompositions for
  gauge amplitudes at tree and loop level}},  {\em Nucl. Phys. B} {\bf 571}
  (2000) 51--70, [\href{http://xxx.lanl.gov/abs/hep-ph/9910563}{{\tt
  hep-ph/9910563}}].

\bibitem{DelDuca:2009au}
V.~Del~Duca, C.~Duhr, and V.~A. Smirnov, {\it {An Analytic Result for the
  Two-Loop Hexagon Wilson Loop in N = 4 SYM}},  {\em JHEP} {\bf 03} (2010) 099,
  [\href{http://xxx.lanl.gov/abs/0911.5332}{{\tt 0911.5332}}].

\bibitem{DelDuca:2010zg}
V.~Del~Duca, C.~Duhr, and V.~A. Smirnov, {\it {The Two-Loop Hexagon Wilson Loop
  in N = 4 SYM}},  {\em JHEP} {\bf 05} (2010) 084,
  [\href{http://xxx.lanl.gov/abs/1003.1702}{{\tt 1003.1702}}].

\bibitem{Korchemsky:1985xj}
G.~P. Korchemsky and A.~V. Radyushkin, {\it {Loop Space Formalism and
  Renormalization Group for the Infrared Asymptotics of \{QCD\}}},  {\em Phys.
  Lett. B} {\bf 171} (1986) 459--467.

\bibitem{Moch:2004pa}
S.~Moch, J.~A.~M. Vermaseren, and A.~Vogt, {\it {The Three loop splitting
  functions in QCD: The Nonsinglet case}},  {\em Nucl. Phys. B} {\bf 688}
  (2004) 101--134, [\href{http://xxx.lanl.gov/abs/hep-ph/0403192}{{\tt
  hep-ph/0403192}}].

\end{thebibliography}\endgroup
\end{document}